\documentstyle[11pt,newpasp,twoside,epsfig]{article}

\begin{document}

\title{Photometric modal discrimination in $\delta$ Scuti and 
$\gamma$~Doradus stars}

\author{Rafael Garrido}
\affil{Instituto de Astrofisica de Andalucia, C.S.I.C., Apdo. 3004, 18080, 
Granada, Spain}

\begin{abstract}

The potential of photometric methods for the identification of $l$, the
degree of spherical surface harmonic of a pulsating star, is investigated
with special emphasis on Str\"omgren photometry applied to $\delta$ Scuti
and $\gamma$ Dor variables.  Limitations of actual model atmospheres
when fine precision is required for the calculations of partial
derivatives and integrals, which depend on limb darkening coefficients,
are discussed. Two methods are discussed to calculate the {\em phase~lags},
the angle between maximum temperature and minimum radius, and $R$, a
parameter which describes departure from adiabaticity of the atmospheres
of these pulsating stars. These quantities appear to be very dependent
on the convection as parametrized by the mixing length theory.  When one
of the methods is applied to the $\gamma$ Dor stars gives {\em phase~lags}
close to $0\deg$, which are $90\deg-180\deg$ out of phase from typical $\delta$
Scuti stars.  Examples are given for some High Amplitude Delta Scuti
Stars (HADS) where the method can be easily applied and gives results
consistent to interpret them as radial ($l$=0) pulsating stars. Other
low amplitude $\delta$ Scuti stars could be oscillating in a non-radial
($l$=1, 2) mode. Multi-band photometry is concluded to be a very powerful
tool for mode identification of $\delta$ Scuti and $\gamma$ Dor stars,
specially with the more accurate photometry that will be achieved in the
near future with the asteroseismological space missions now in progress.

\end{abstract}

\keywords{Pulsating stars, photometry, $\delta$ Scuti stars, $\gamma$ 
Doradus stars, convection, stellar atmospheres, pulsation theory}

\section{Introduction}

In recent years the development of photometric multi-site
campaigns organized by different teams: Delta Scuti
Network {\em (http://dsn.astro.univie.ac.at)}, STEPHI {\em
(http://dasgal.obspmfr/stephi/)} and STACC (Frandsen et al. 1996) for
studying multi-periodic $\delta$ Scuti stars has increased the number of
detected frequencies up to a level at which we could think that direct
fitting to a theoretical model could be relevant to test the modelization
and finally to do real asteroseismology of these stars. See for instance
Breger et al. (1999) for FG Vir and Pamyatnykh et al. (1998) for XX Pyx.
However the number of theoretically excited and photometrically visible
radial and non-radial periods in a $\delta$ Scuti is so large that it is
impossible to match all theoretical frequencies to the observed ones.
For this reason a method to identify modes is needed.  One can think
that the rotational splitting is the most simple method and to look for
singlets (radial, $l=0$), triplets (dipole, $l=1$), etc., in the
frequency spectrum. However the overlapping and non-linear interactions
among them destroy the expected regular pattern as demonstrated in Breger
et al (1999). In any case some residual regularity can be expected
and used as an indication of the degree of the spherical harmonic $l$
of the observed modes, as explained there.  Therefore identification
of at least a few frequencies becomes necessary to attempt to make
asteroseismological techniques available for $\delta$ Scuti stars.

Another approach, better explained in other reviews of this conference,
consists of the use of the line profile variations to model the
surface velocity field and then deduce the corresponding spherical
harmonic. Technical details are given elsewhere, Mantegazza in these
proceedings and Aerts these proceedings as well, but I would like to
note here the difficulty to get sufficient spectroscopic measurements,
since both high S/N ratio and time coverage to resolve frequencies are
required, so that  medium size telescopes are needed. In fact it is
quite difficult to organize multi-site spectroscopic campaigns for many
days, but easier when the same is done for small photometric telescopes.
However the enormous advantage of the spectroscopy is that one can obtain
information not only on the $l$-values but also on the azimuthal order of
the spherical harmonic $m$-values, which are not detectable from purely
photometric observations.

The main question we want to address here is whether multi-band photometry 
alone can be useful to discriminate $l$-values or not.  

Based on the linearization made by Dziembowski (1977) of the bolometric
magnitude variations exhibited by a star undergoing non-radial
oscillations, Balona \& Stobie (1979a, 1979b) formulated an analytic
expression which will be used throughout this review in the form given
by Watson (1988).  The equation, as we will see later on, contains
evaluations of flux derivatives and of integrals over limb darkening
coefficients. This requires a well-behaved model atmospheres not only
describing mean values, from which we can deduce physical parameters,
but also their variations as far as temperature and/or gravity varies.

I will start by introducing the linearized equation and the physical
conditions required for applicability, then I will discus the
limitations of the available model atmospheres, basically Kurucz models
and modifications, concerning partial derivatives and limb darkening
coefficients. As we will see, there are two unknown quantities in the
formula, the {\em phase~lag} and a parameter related with departures from
adiabaticity, which deserves a detailed discussion given in section
3. The next section is devoted to the application to real data mainly
through the use of Str\"omgren photometry. Finally in the last section,
I will present the expected possibilities of the method in the context
of future space missions.

\section{The linearized equation}

Thermal time scales for $\delta$ Scuti stars are at least an order of
magnitude larger than the shortest observed period. Furthermore the
relative radius variations, even for the High Amplitude $\delta$ Scuti
Stars (HADS), are so small that second order terms can be neglected.

Under such conditions the linearized equation governing the photometric
variation of a pulsating star, $\delta x(t)$, can be expressed, following
the formulation given in Watson (1988), as:

\begin{eqnarray}
\delta x(t) & = & -1.086 \,\epsilon P_{l,|m|}(\cos(\theta)) \times \nonumber\\
            &   & [(T_1 + T_2) \cos(\omega t + \Psi^T) + 
	    (T_3 + T_4 + T_5) \cos(\omega t)]
\end{eqnarray}
where the variables have the following definitions:
\begin{itemize}

\item  the $x$'s stand for the different photometric bands. Here I
         will use $u, v, b$ and $y$ of the Str\"omgren photometric system.

\item  $\epsilon \ll 1$ is an arbitrary small quantity.

\item  $P_{{l,|m|}}(\cos(\theta))$ is the associated Legendre polynomial
         of order ${l,m}$ to the pulsation frequency $\omega$,
         oriented with the angle $\theta$, i. e. the inclination of the
         stellar pulsation axis to the observer.

\item  $\Psi^T$ is the {\em Phase~Lag}, i.e. the angle between maximum
        temperature and minimum radius, hence $\Psi^T = \pi$ in perfect
        adiabatic conditions.

\item  The $T_1 \ldots T_5$ can be written as follows:

\begin{eqnarray} 
 T_1 & = & b_{l,x}B \frac{\partial x(t)}{\partial \log T}\;,
 \\ & \nonumber & \\ 
 T_2 & = & \frac{B}{2.3026} \frac{\partial b_{l,x} }{\partial \log T}\;,
 \\ \nonumber & & \\ 
 T_3 & = & b_{l,x}(2+l)(1-l)\;,
 \\  \nonumber & & \\ 
 T_4 & = & -b_{l,x}p^*C \frac{\partial x(t)}{\partial \log {\sl g}}\;,
 \\  \nonumber & & \\ 
 T_5 & = & -\frac{p^*C}{2.3026} \frac{\partial b_{l,x} }{\partial \log {\sl g}}
 \\ \nonumber & &
\end{eqnarray}

\end{itemize}

 where $b_{l,x}$ are the weighted limb darkening integrals defined as:

\begin{eqnarray} 
  b_{l,x} & = & \int_0^1 h_{\lambda} (\mu)\mu P_l(\mu)d\mu 
\end{eqnarray}

where, using a quadratic limb darkening law,

\begin{eqnarray} 
 h_{\lambda} (\mu) & = & \chi_{_0} + \chi_{_1}\mu + \chi_{_2}\mu^2
\end{eqnarray}

Partial derivatives and limb darkening coefficients have to be tabulated from 
theoretical model atmospheres.
$p^*$ is a measure of the variation of the pressure when a gravity variation 
occurs, evaluated at an optical layer where we observe the continuum flux. It 
is then 

\begin{eqnarray}
 p^* = \left(\frac{\partial {\sl \log g}}{\partial {\log p}}\right)_{_{\tau=1}}
\end{eqnarray}
From model atmospheres and in the $\delta$ Scuti regime this value can be
considered as a constant equal to approximately 1.4.

$B$ and $C$ are related by the equation:
\begin{eqnarray}
B = R \left(1 - 1/\Gamma_2\right) C
\end{eqnarray}
where $C$ is given by:
\begin{eqnarray}
C = (4 + 1/\alpha_{_H}) - l(l + 1)\alpha_{_H}
\end{eqnarray}
and 
\begin{eqnarray}
\alpha_{_H} = G\rho_{_{\odot}}Q^2/3\pi.
\end{eqnarray}

$Q$ is calibrated, see for instance Breger et al. (1993), using the 
following equation:
\begin{eqnarray}
\log Q = -6.456 + \log P + 0.5 \log {\sl g} + 0.1M_{\rm bol} + \log T_{\rm eff}.
\end{eqnarray}
Standard photometric indices are then used to calibrate gravity, absolute 
magnitude and temperature and, by knowing the pulsation period, we can obtain  
the observational value for the pulsation constant $Q$.

$R$ is introduced as a free parameter to estimate deviations from 
adiabaticity, since for an adiabatic atmosphere:
\begin{eqnarray}
B = (1 - 1/\Gamma_2) C
\end{eqnarray}
then $R$ must be unity in that case and physical values for $R$ must be
between $0$ and $1$. The adiabatic exponent $\Gamma_2$ can show dramatic
changes in the atmosphere of a pulsating star but in our case we assume a
constant value of $5/3$. So any change of this parameter will change the
value of $R$, which will remain as an unknown parameter in equation (1).

The general procedure described here to obtain $l$-values from observed
photometric indices is explained in the flow chart of Fig~1. It is based
on the use of model atmospheres to calculate not only global physical
quantities for a given star but also variations of the photometric indices
with respect to temperature and gravity, together with the above defined
integral of the limb darkening coefficients. However, and in order to
know the corresponding $l$-value, we need to have an estimation of the
two unknown variables $\Psi^T$ and $R$. Before doing that in the next
section we will discuss more in detail the precision of these quantities
when model atmospheres from Kurucz (1993), i.e. ATLAS9 code, are used.

Starting by the observables $x$, the calibrated photometric indices,
$P$, the pulsation period and the time variations of the photometric
indices $x$, $\delta x$ , which include phases and amplitudes, we are
able to obtain for the $\delta$ Scuti and related stars the following:

\begin{enumerate}
\item Effective temperatures, gravity and metallicity from model
            atmospheres

\item Limb darkening coefficients and partial derivatives, also
            from model atmospheres

\item Pulsation constant from equation (13) and the previously calibrated 
            quantities
\end{enumerate}

 There are three quantities which remain unknown from equation (1): the 
{\em phase~lag}, $\Psi^T$, the adiabaticity parameter, $R$, and the $l$-value 
we are searching for.
 At this point two strategies exist:

\begin{itemize}

\item to make an assumption with physical sense about $\Psi^T$
      and $R$.  Usually $R$ is selected in the range $0.25<R<1$ only
      from theoretical considerations whereas $\Psi^T$ is selected in
      the range of $90\deg<\Psi^T<135\deg$ from spectroscopic observations of
      $\delta$ Scuti stars. One can draw then the corresponding ``regions
      of interest'' for the different $l$-values. A comparison of the
      phase differences {\it vs.} amplitude ratios with observations
      is the approach followed by different authors (Watson, 1988 and
      Garrido et al., 1990).

\item to estimate $\Psi^T$ and $R$ either from theoretical calculations
      (Balona and Evers, 1999) or by photometric observations as explained
      in detail in Garrido et al. (1990).

\end{itemize}

In both cases the final estimation of the $l$-value is made by
selecting, through a least squares procedure, the minimum distance
between predictions and observations as we will see in section 5.2.

\begin{figure}
\plotfiddle{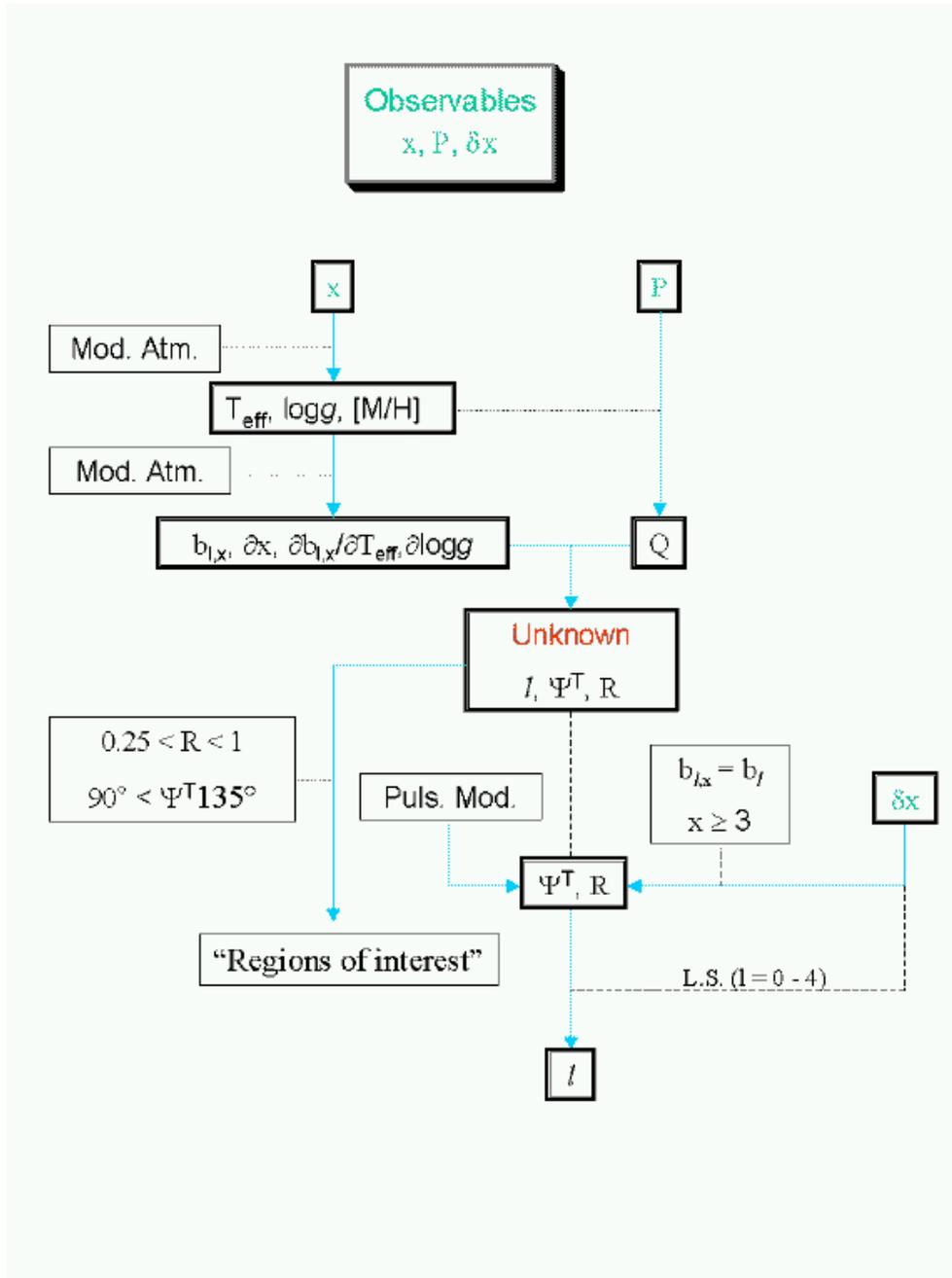}{15cm}{0}{68}{68}{-205}{-40}
\caption{Flow chart describing the main parts of this review. See text for 
an explanation of the different symbols.
\label{fig-1}
}
\end{figure}

Following the flow chart in Fig~1 we will next discuss in detail a practical 
application by using Str\"omgren data and Kurucz models.

 \section{Kurucz model atmospheres}

 \subsection{Physical calibration}

The Str\"omgren system is a photometric system composed of 4 filters
of an intermediate width centered at $u: 3500$ \AA, $v: 4100$ \AA, $b:
4670$ \AA\ and $y: 5470$ \AA. They are very well adapted to derive physical
parameters for the spectral types we are dealing with: $\delta$ Scuti
and related stars.

A direct comparison with the Kurucz models, without any correction, 
of the standard stars as defined in Smalley and Kupka (1997) gives  
the following results for the 6 primary standard stars:\\

 Standard ATLAS9~~~~~~~~~~~~~~$T_{\rm eff}({\rm model}) - T_{\rm eff}({\rm
 observed}) = 88 \pm
 120$ K    \\

 Standard ATLAS9~~~~~~~~~~~~~~~$\log\,{\sl g}({\rm model}) - \log\,{\sl
 g}({\rm observed}) = -0.23 \pm 0.18$\\

\noindent and for the 15 IRFM standard stars:\\
\newpage

 Standard ATLAS9~~~~~~~~~~~~~~$T_{\rm eff}({\rm model}) - T_{\rm eff}({\rm observed}) = 164 \pm 
131$ K    \\

 Standard ATLAS9~~~~~~~~~~~~~~$\log\,{\sl g}({\rm model}) - \log\,{\sl g}({\rm observed}) = -0.30 \pm 0.34$\\

Using the modification indicated by Smalley and Kupka (1997) which 
consists basically in the use a new treatment of the turbulent convection 
developed by Canuto and Mazzitelli (1991, 1992), one finds for the 6 primary 
standard stars:\\

 Modified ATLAS9~~~~~~~~~~~~~~$T_{\rm eff}({\rm model}) - T_{\rm eff}({\rm observed}) = 75 \pm 
115$~K    \\

 Modified ATLAS9~~~~~~~~~~~~~~$\log\,{\sl g}({\rm model}) - \log\,{\sl g}({\rm observed}) = -0.02 \pm 0.05$\\

\noindent
and for the 15 IRFM standards stars:\\ 

 Modified ATLAS9~~~~~~~~~~~~~~$T_{\rm eff}({\rm model}) - T_{\rm eff}({\rm observed}) = 109 \pm 
118$~K    \\

 Modified ATLAS9~~~~~~~~~~~~~~$\log\,{\sl g}({\rm model}) - \log\,{\sl g}({\rm observed}) = -0.10 \pm 0.12$\\

The small difference is due to different assumptions for the
physical atmospheric values for Vega: Smalley and Kupka (1997) using
spectrophotometry from Hayes (1985), obtain $T_{\rm eff}$=9550 K,
$\log\,g$=3.95, $[M/H]$=-0.5 and a micro-turbulence of $2.0$~km~s$^{-1}$
whereas Kurucz uses that of Hayes and Latham (1975) obtaining: $T_{\rm
eff}$=9400 K, $\log\,{\sl g}$=3.90, [M/H]=-0.5 and a micro-turbulence of
0.0~km~s$^{-1}$.  However the Smalley and Kupka (1997) calibration
seems to be better concerning gravities, where they obtain a lower
dispersion. In any case, and for our purposes, we can establish that the
best temperature calibration we can have is around 100 K in error, and
the best gravity is good within around 0.1 dex in $\log\,g$. Metallicity
as measured with the index $m_0$ can be calibrated using the procedure
given in Smalley (1993) to which we refer for a detailed discussion. In
any case the discrepancy shown by the models introduces an uncertainty
of 0.5 in [M/H] which is not dramatic as compared to other uncertainties
related with the derivatives as we will see in the following subsection.

\subsection{Partial derivatives}

Once we have a calibrated value for $T_{\rm eff}$, $\log\,{\sl g}$ and
[M/H] we can select a grid and perform the needed partial derivatives
with respect to $T_{\rm eff}$ and $\log\,{\sl g}$ in equation (1). These
grids are sampled at 500 K in $T_{\rm eff}$ and 0.5 dex in $\log\,{\sl
g}$. Cubic splines subroutines have been used to calculate them.

\begin{figure}[t]
\plotfiddle{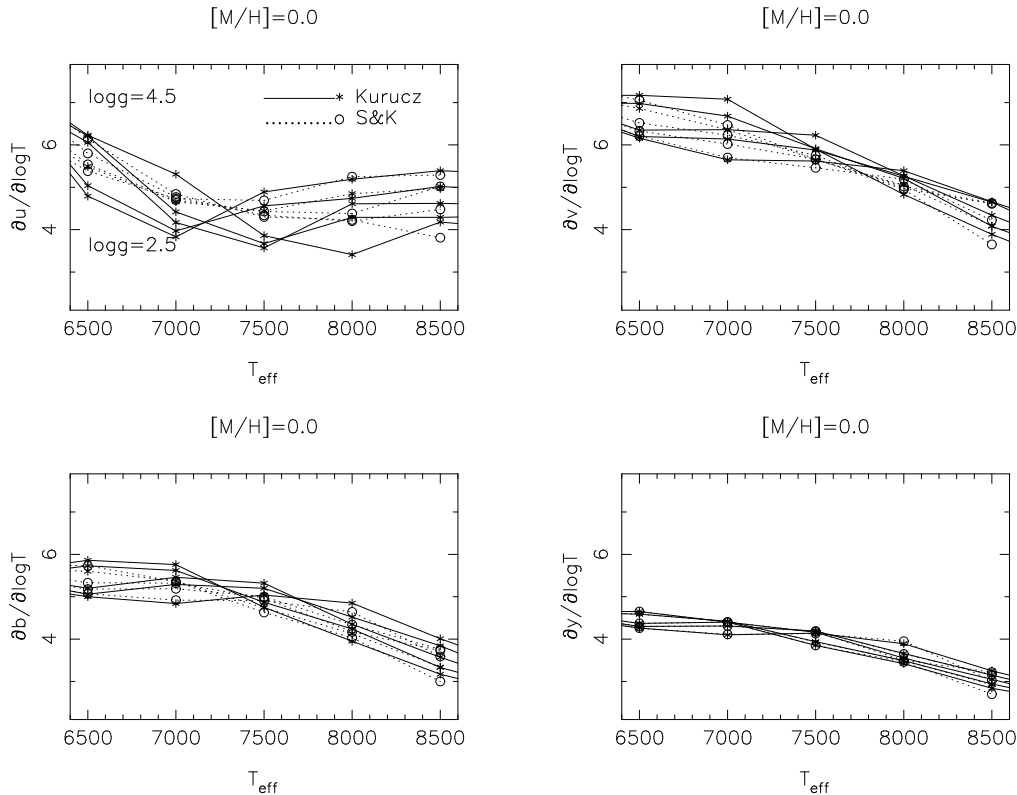}{10cm}{-90}{55}{55}{-210}{315}
\caption{Partial derivatives of the Str\"omgren $uvby$ fluxes, in
magnitudes, with respect to the temperature in the range  6500~K $\leq
T_{\rm eff} \leq 8500$~K, 2.5 $\leq \log\,{\sl g} \leq 4.5$~and~$[M/H]=0$.
\label{fig-2}
}
\end{figure}

\begin{figure}[t]
\plotfiddle{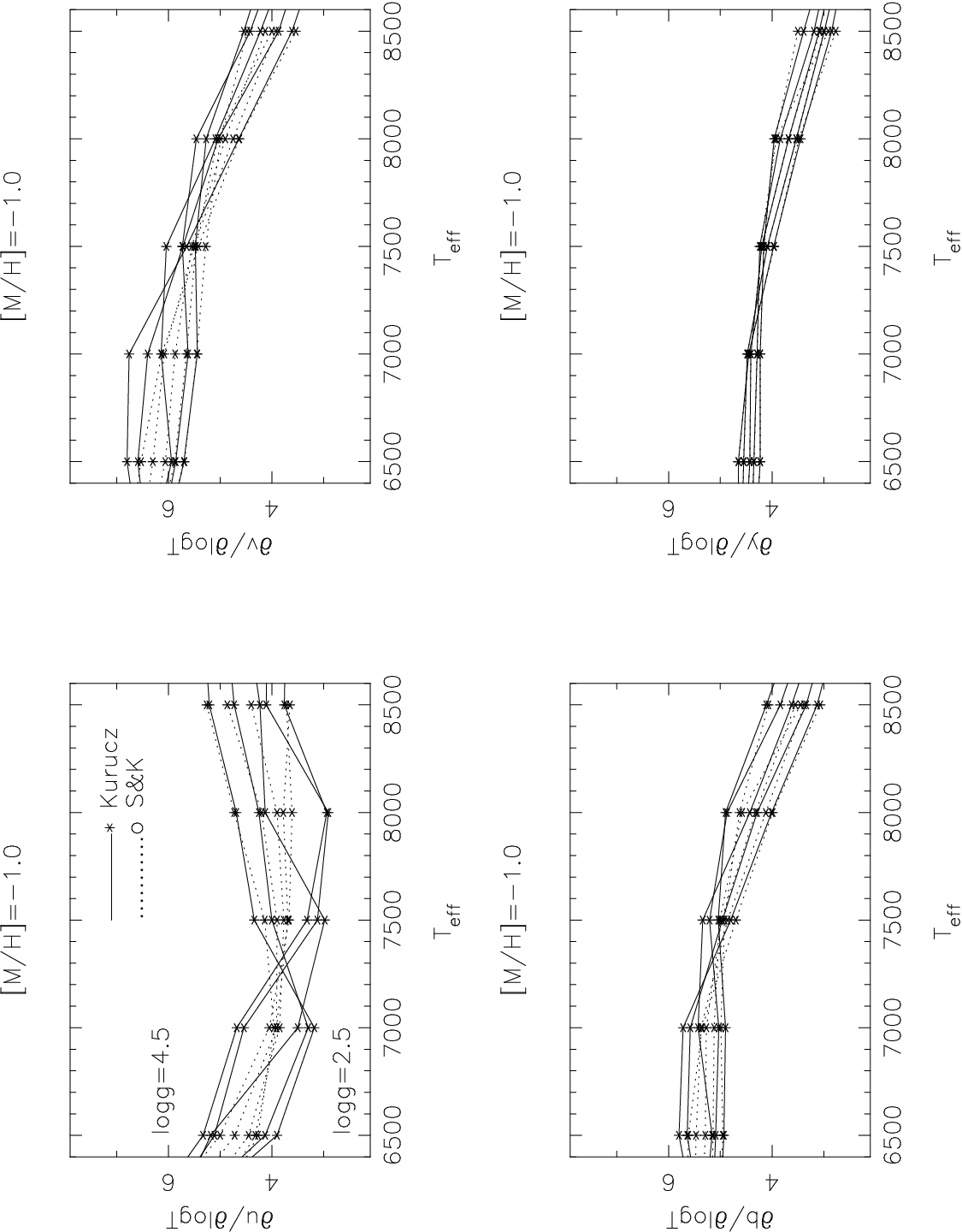}{10cm}{-90}{55}{55}{-210}{315}
\caption{Partial derivatives of the Str\"omgren $uvby$ fluxes, in
magnitudes, with respect to the temperature in the range  6500~K $\leq
T_{\rm eff} \leq 8500$~K, $2.5 \leq \log\,{\sl g} \leq 4.5$~and~$[M/H]=-1$.
\label{fig-3}
}
\end{figure}

In Fig~2 the partial derivatives of the original Pop I $[M/H]=0$
Kurucz models with respect to $T_{\rm eff}$ are plotted, models are
with overshooting and $[l/H]=1.25$. They are marked with full lines,
and those of the Kurucz models modified by Smalley and Kupka (1997;
hereafter S\&K), are marked with dots.  The main difference is that S\&K
models are calculated with a different convection treatment, as explained
in detail in their paper. Ranges in $T_{\rm eff}$ and $\log\,{\sl g}$ are
those for $\delta$ Scuti and related stars. They are clearly decreasing
functions with increasing temperature for $b, v$ and $y$ filters and show
a not very pronounced minimum in the ultraviolet $u$ band.  There is also
a general trend in the sense that blue bands are noisier than red ones,
independent of the models used. Models are essentially the same for the
$y$ visible band; however the modified S\&K models are smoother than
original Kurucz ones. For the $u$ band this effect is magnified but, in
any case, at temperatures higher than 7500 K, even S\&K models start to
show some small discontinuities. Uncertainties in these derivatives can
reach up to $20\%$ in mag/$\log T$ units, depending on the model and/or the
($T_{\rm eff}$, $\log\,{\sl g}$) regime that we are selecting.

There is basically no differences when a non solar metallicity is used
(see Fig~3) but, as expected, partial derivatives of the blue bands are
lower than for solar metallicity: a metal deficiency of the order of
$[M/H]=-1$ makes a significant contribution to the flux depletion at
these wavelengths and consequently a lower dependence.

Concerning the partial derivatives with respect to $\log\,{\sl g}$
the behavior is very similar in the sense that they are noisier for
blue colors and for the original Kurucz models, as can be seen in
Fig~4. Furthermore they are all almost constant over the ($T_{\rm eff}$,
$\log\,{\sl g}$) range shown in the figure.  However the amplitude of
variations of these derivatives for the original Kurucz models can reach
values of the order of $100\%$ for some temperatures and gravities! Here
again the S\&K modified models behave better, always in the sense that
these are smoother than the original Kurucz ones.

One can think that these uncertainties in the partial derivatives would make 
the method useless but we will see in the next section that the relative 
importance of each term in equation (1) enables some possibilities for 
discriminating $l$-values depending on the filters and physical conditions 
even considering these uncertainties. 
    
\begin{figure}[t]
\plotfiddle{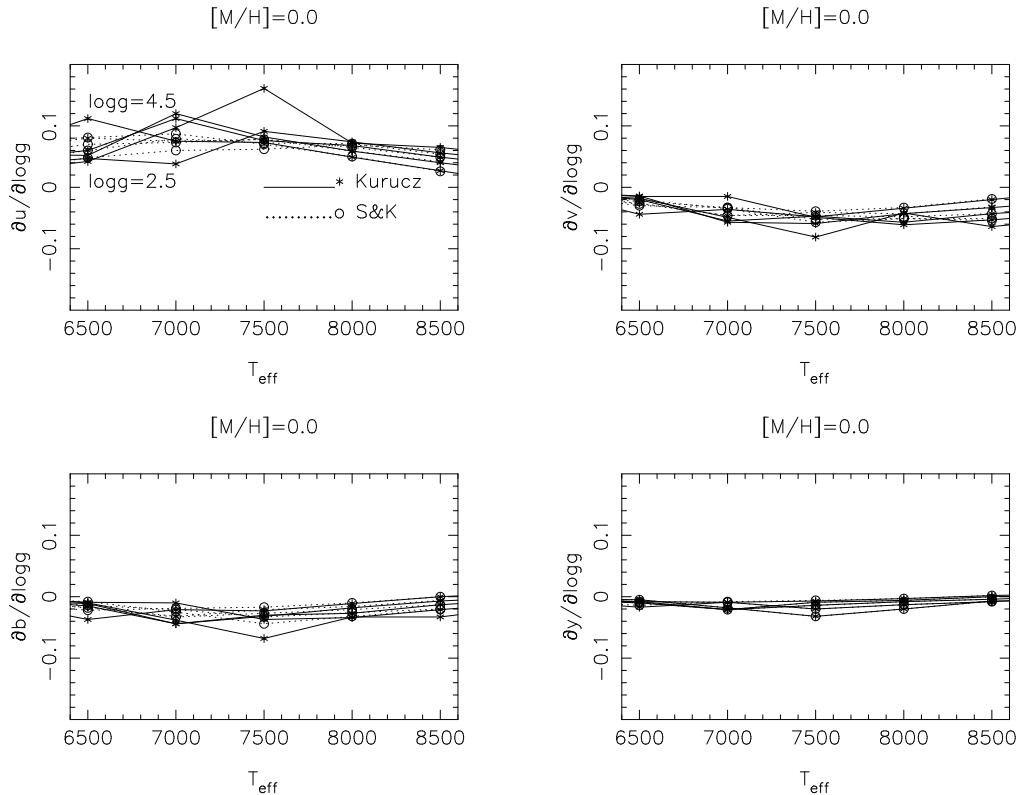}{10cm}{-90}{55}{55}{-210}{315}
\caption{Partial derivatives of the Str\"omgren $uvby$ fluxes, in
magnitudes, with respect to the log of gravity in the range  6500~K
$\leq T_{\rm eff} \leq 8500$~K, 2.5 $\leq \log\,{\sl g} \leq 4.5$~and~$[M/H]=0$.
\label{fig-4}
}
\end{figure}

\subsection{Limb darkening integrals and derivatives}

Limb darkening integrals defined in (7) with quadratic limb darkening
laws for the Str\"omgren photometric bands $uvby$ have been calculated
using the coefficients given in Table II, page~262, of Watson (1988).
Here the step in temperature is 250~K and I only show values for 3
different gravities, 3.5, 4 and 4.5 dex.

Integrals corresponding to $l=1, 2, 3$~and$~4$ for a solar metallicity 
atmosphere are shown in Fig~5, 6, 7 and 8. Radial values, $l=0$ are 
exactly unity from the definition given in (7). They are slowly decreasing 
functions as temperature increases and with a not very well defined trend  
for gravities ranging from 3.5 to 4.5 dex. It is also to be noted that there  
 is a slight wavelength dependence which becomes more important for high 
$l$-values. This will be crucial for the method developed originally by 
Garrido et al. (1990) and explained in section~5.

\begin{figure}[p]
\plotfiddle{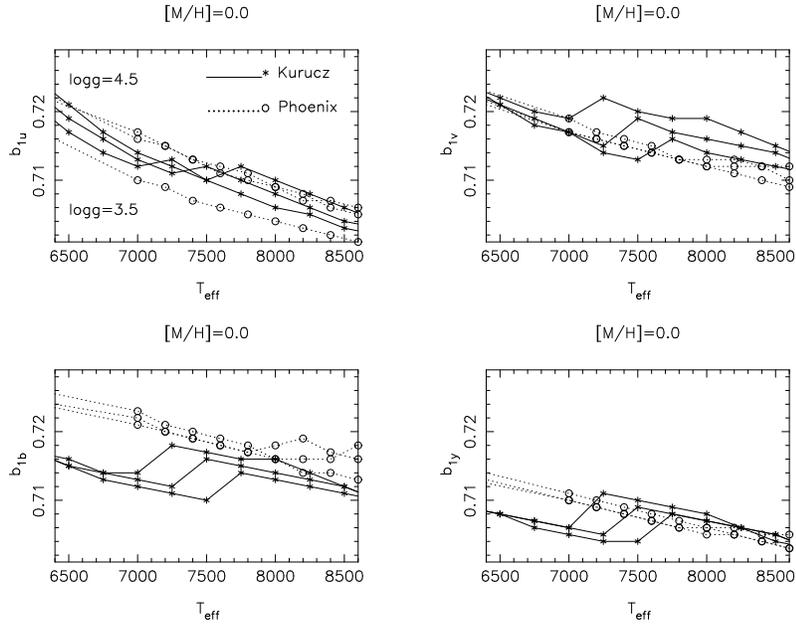}{8cm}{-90}{43}{43}{-165}{250}
\caption{Limb darkening integrals for the Str\"omgren $uvby$ bands 
 in the range  6500~K $\leq T_{\rm eff} \leq 8500$~K, 3.5 
$ \leq \log\,{\sl g} \leq 4.5$~and~$[M/H]=0$ for $l=1$.
\label{fig-5}
}
\end{figure}

\begin{figure}[p]
\plotfiddle{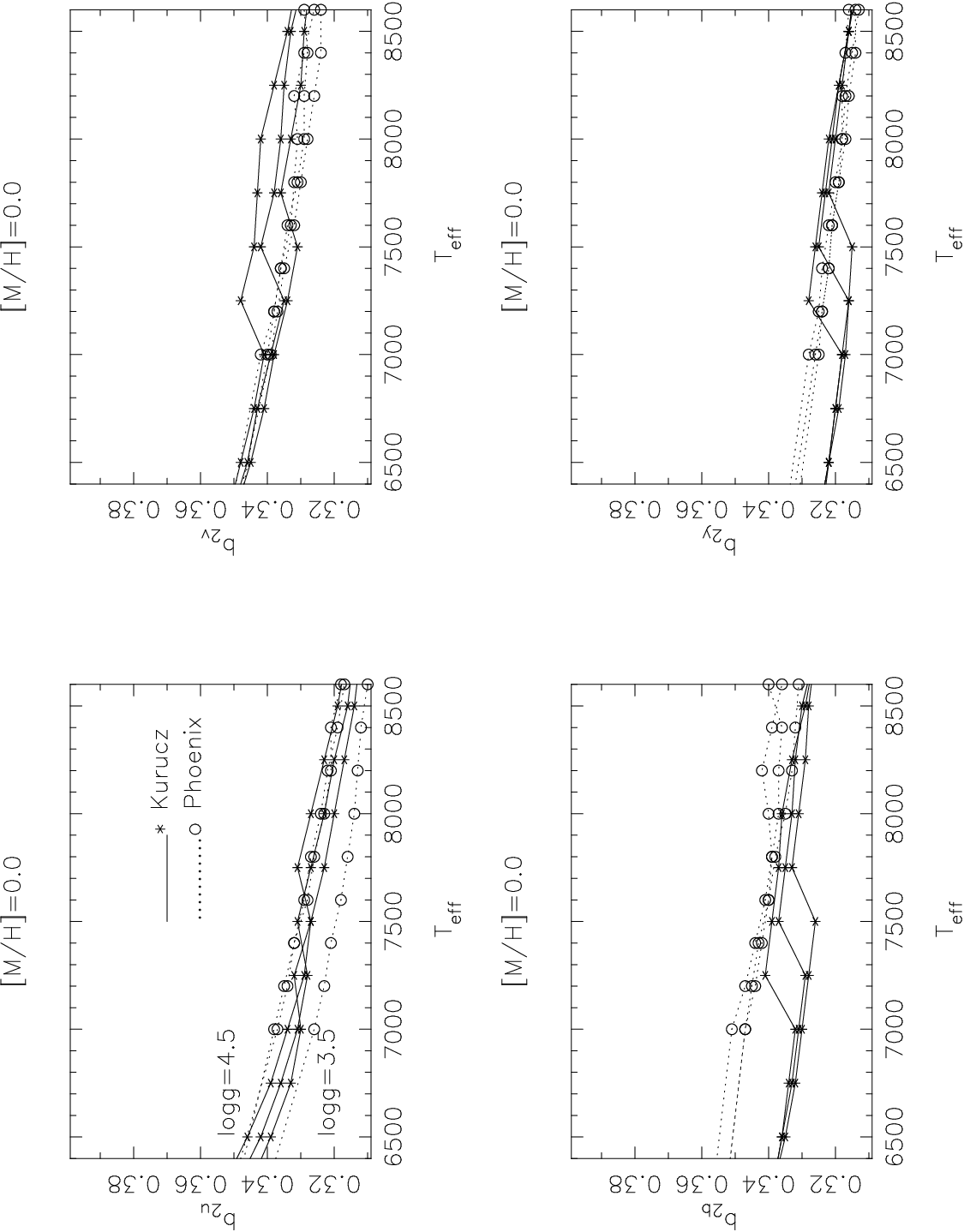}{8cm}{-90}{43}{43}{-165}{250}
\caption{Limb darkening integrals for the Str\"omgren $uvby$ bands 
 in the range 6500~K $\leq T_{\rm eff} \leq 8500$~K, 3.5 
$ \leq \log\,{\sl g} \leq 4.5$~and$~[M/H]=0$ for $l=2$.
\label{fig-6}
}
\end{figure}

\begin{figure}
\plotfiddle{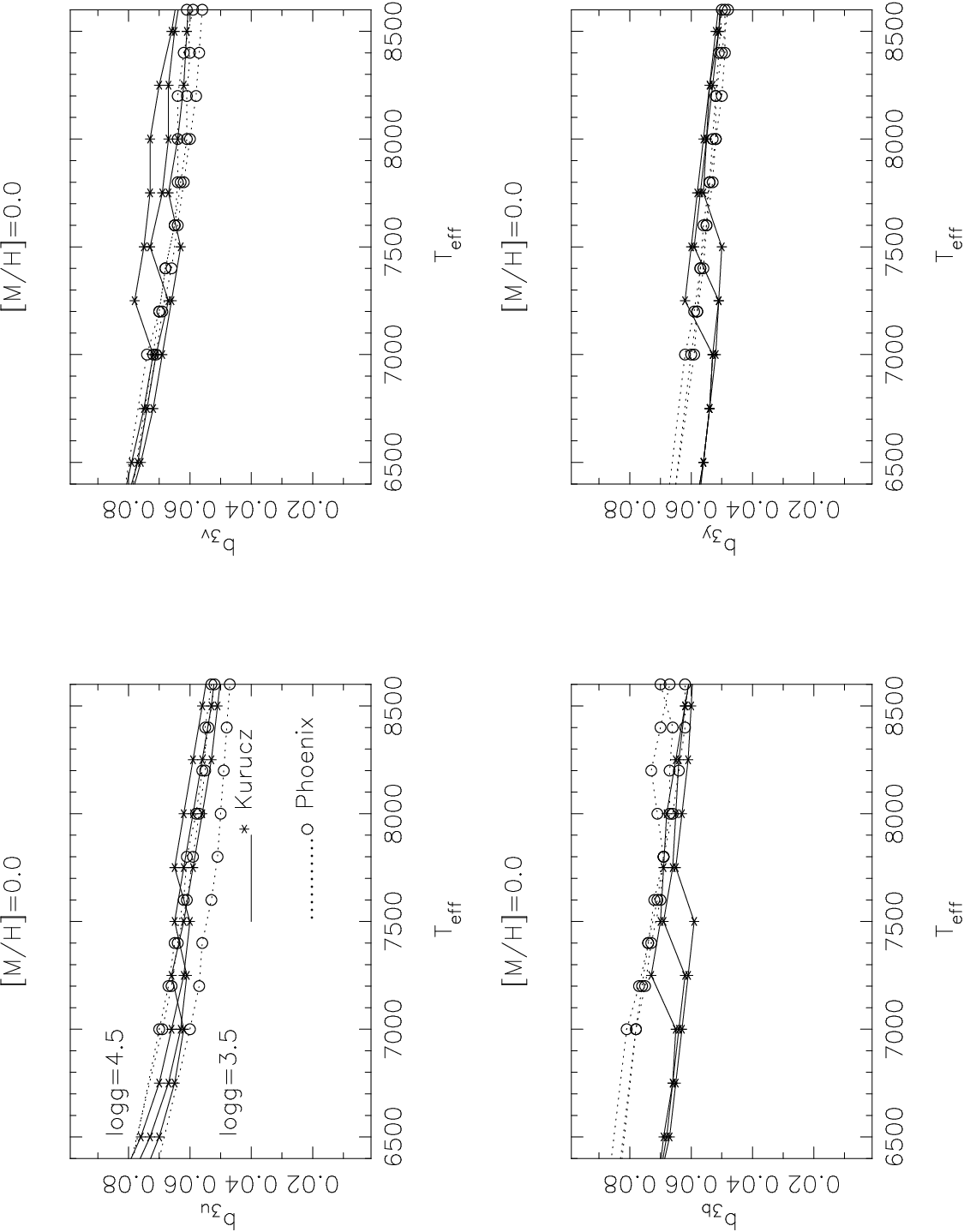}{8cm}{-90}{43}{43}{-165}{250}
\caption{Limb darkening integrals for the Str\"omgren $uvby$ bands 
in the range  6500~K $\leq T_{\rm eff} \leq 8500$~K, 3.5 
$ \leq \log\,{\sl g} \leq 4.5$~and~$[M/H=0]$ for $l=3$.
\label{fig-7}
}
\end{figure}

\begin{figure}
\plotfiddle{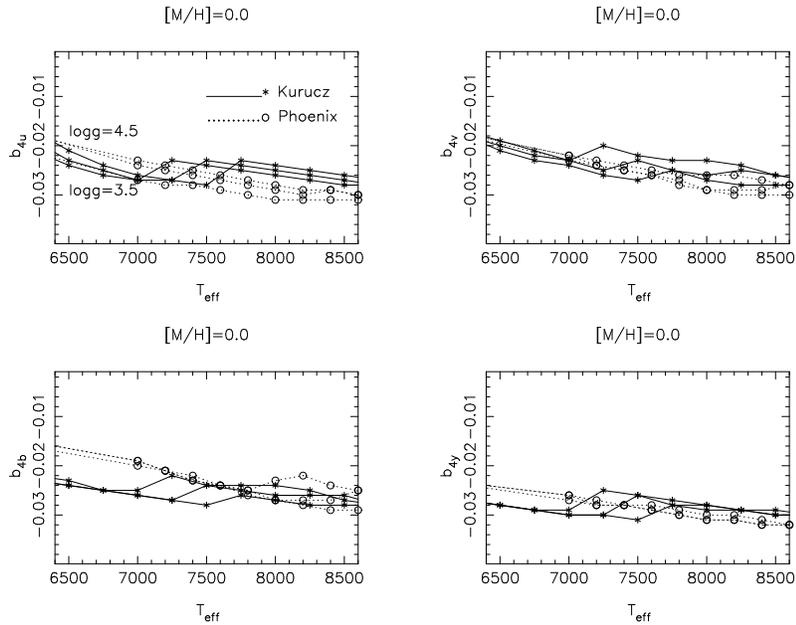}{8cm}{-90}{43}{43}{-165}{250}
\caption{Limb darkening integrals for the Str\"omgren $uvby$ bands 
in the range 6500~K $\leq T_{\rm eff} \leq 8500$~K, 3.5 
$\leq \log\,{\sl g} \leq 4.5 $~and$~[M/H]=0$ for $l=4$.
\label{fig-8}
}
\end{figure}

One important feature of these plots are the discontinuities appearing
at around 7000--7500 K in the original Kurucz models. This effect,
which can be also seen in Fig~5 of the Watson (1988) review, is also
independent of the wavelength and describes some inconsistency in these
models. I selected, for comparison purposes, the new models developed
in the PHOENIX code (see Hauschildt 1992, 1993; Hauschildt and Baron
1995; Baron et al. 1996 for a detailed description). In these models
no discontinuity is seen although some inconsistencies still persist,
in particular concerning the gravity variation, (see for instance the
difference between $\log\,{\sl g}=3.5$ and the other two derivatives at
$\log\,{\sl g}=4, 4.5$ in the $u$ band for $l=1, 2$ or the small deviations
at high temperatures depending on the gravity).  Discontinuities appearing
in the Kurucz models at 7000--7500 K seem to be produced by the arbitrary
suppression of overshooting in these models. In any case PHOENIX does
not use overshooting and the effect is not present.

The effect is extraordinarily enhanced when we calculate derivatives,
shown in Fig~9 and 10, which have also to be calculated as indicated in
equation (1). It is important to note here that the inclusion of the limb
darkening integrals variations into that equation is irrelevant if we
use the standard original Kurucz models, since uncertainties are of the
same order as the derivatives.  The situation is improved when one use
the PHOENIX models but even so there are some regions (high temperature
regions, double value for the partial derivatives in the $u$ band for
different gravities, \ldots) where these uncertainties still remain.

\begin{figure}[p]
\plotfiddle{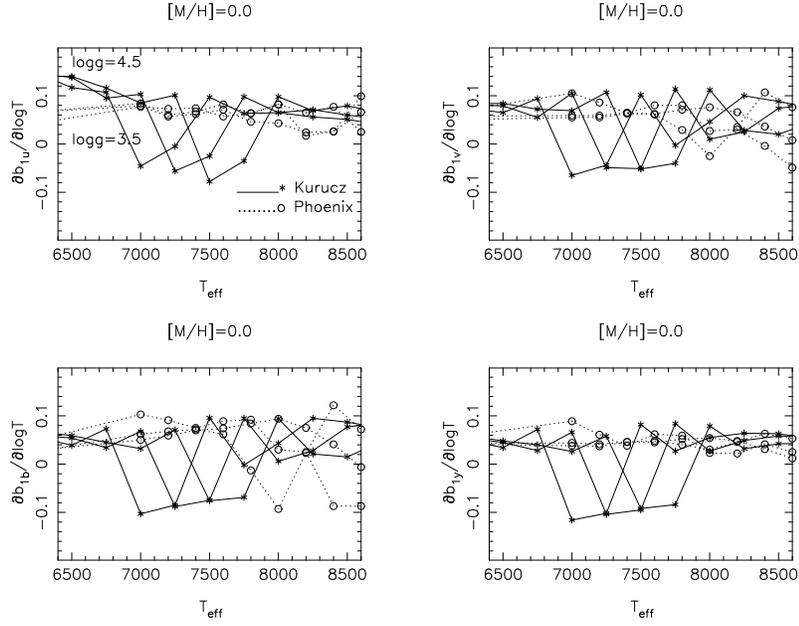}{8cm}{-90}{43}{43}{-165}{250}
\caption{Partial derivatives of the limb darkening integrals for the
Str\"omgren $uvby$ bands, with respect to the log of temperature, in the
range  6500~K $\leq T_{\rm eff} \leq$ 8500~K, 2.5 dex $\leq \log\,{\sl g}
\leq 4.5$ dex and $[M/H]$=0.
\label{fig-9}
}
\end{figure}

\begin{figure}[p]
\plotfiddle{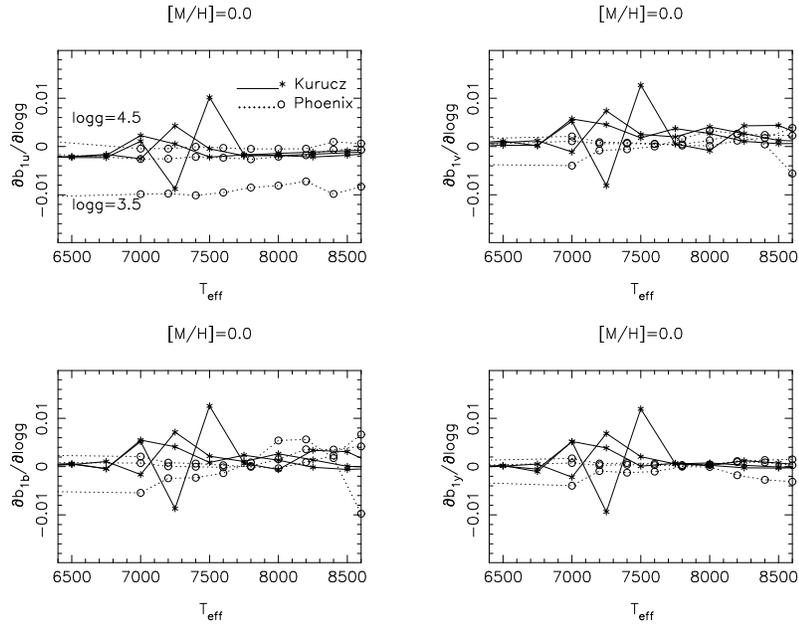}{8cm}{-90}{43}{43}{-165}{250}
\caption{Partial derivatives of the limb darkening integrals for the 
Str\"omgren $uvby$ bands, with respect to 
the log of gravity in the range  6500~K $\leq T_{\rm eff} \leq$ 8500~K, 2.5 
dex $\leq \log\,{\sl g} \leq$ 4.5 dex and $[M/H]$=0.
\label{fig-10}
}
\end{figure}
 
At first sight these large inconsistencies in the derivatives and in
the limb darkening integrals and their derivatives seem to introduce
large uncertainties in equation (1), but the situation is improved when
we realize the different contributions from different terms in that
equation. In Fig~11 I show these different contributions for a typical
$\delta$ Scuti regime: ($\log T_{\rm eff}=7500$~K, $\log\,{\sl g}=4,
[M/H]=0$), and with a mean value for the adiabaticity parameter, i.e.
$R$=0.5. The different filters and filter combinations are the classical
ones for the Str\"omgren photometric system, i.e. $(b-y)$ as a temperature
indicator and $c$ as a luminosity indicator (at least for stars close
enough to the Main Sequence).

As expected, the main contribution to any band and $l$-value comes
from temperature variations. Geometry variations, given by the term
defined in (4), is the second important parameter for low $l$-values
but for $l$=3 and 4 can become as large as, or even larger than,
the temperature term defined in (2). It is also to be noted the large
effect of the gravity variations in the color index $c$, as properly
defined by the Str\"omgren photometry. The basic idea is to use the
geometry dependence, through the term defined in (4), for $l$=0, 1 and
2 because of the changing sign of this term when passing from $l$=0 to
$l$=2 and the null contribution for $l$=1. So the main utility is not
basically changed by the uncertainties of the other terms in equation
(1) so allowing, for the lowest $l$-values, a proper discrimination as
we will see later on. In any case the term defined by (6) is always
very small and the term defined by (3), no very well known neither,
is only important for $l$=4 when the smearing out effects along the
stellar surface make the photometric method inapplicable.

\begin{figure}
\plotfiddle{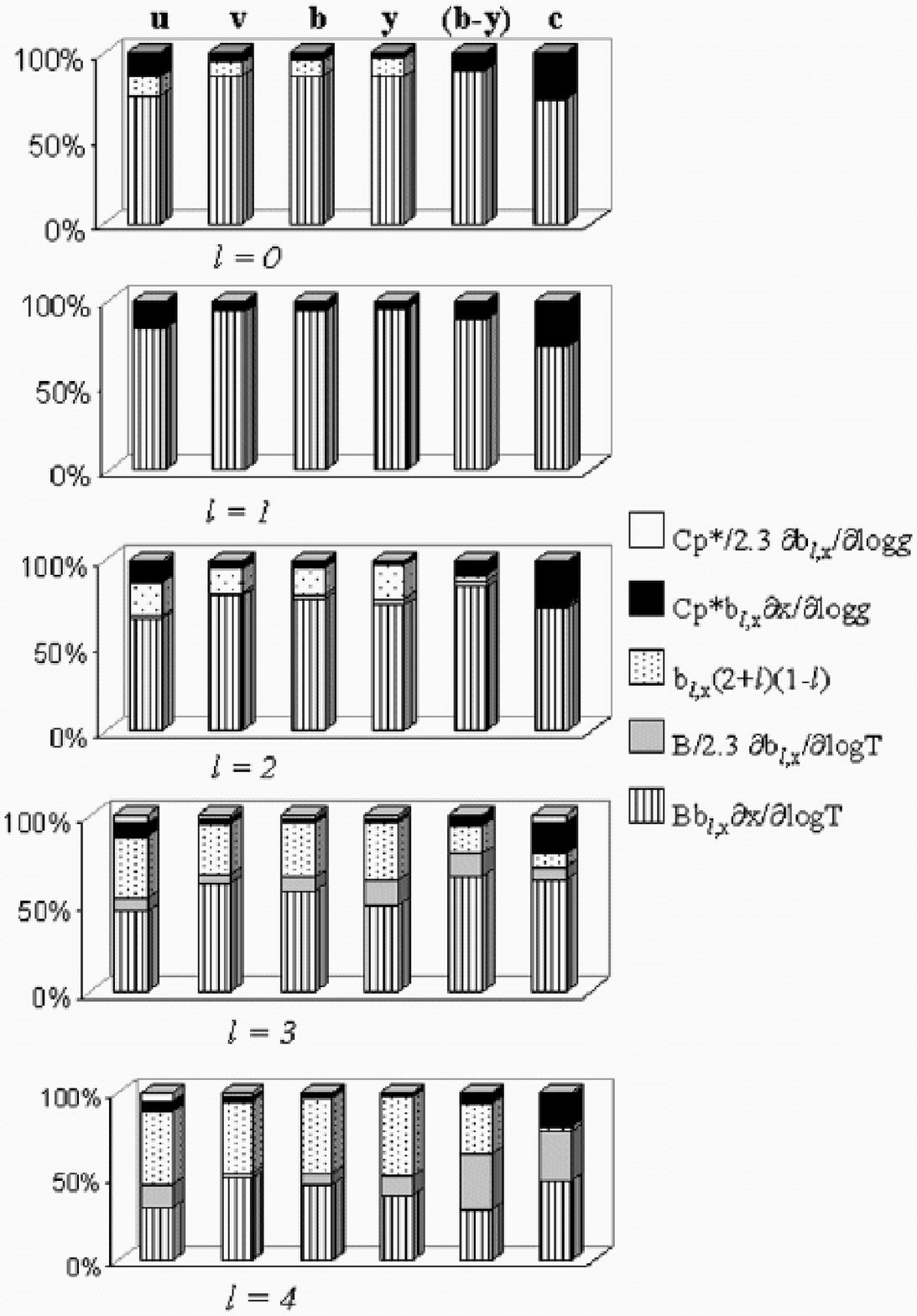}{17cm}{0}{70}{70}{-180}{-20}
\caption{Percentage of contribution to the different Str\"omgren bands and
colors for a $\delta$ Scuti regime ($\log T_{\rm eff}=7500$~K,~$\log\,{\sl
g}=4,~[M/H]=0$) and $R$=0.5.  
\label{fig-11} } 
\end{figure}

\section{``Regions of interest''}

Once $T_{\rm eff}, \log\,{\sl g}, [M/H]$, derivatives and limb darkening
have been calculated for a given star two unknown quantities in equation
(1), $(\Psi^T, R)$, still remain in order to obtain the quantity we are
looking for: $l$, the degree of the spherical surface harmonic. $(\Psi^T$
is not a very well known quantity but it can be estimated from
simultaneous photometric and radial velocities observations.  Typical
values for some $\delta$ Scuti given in Breger et al. (1976) range from
$90\deg$ to $140\deg$ but {\em phase~lags} for $\gamma$ Dor stars are
completely unknown. On the other hand the adiabaticity parameter $R$ can
not be known from pure observations and one must assume some reasonable
value for it which, according to equation (11), is restricted to values
between 0 and 1.

 In what follows I will adopt a constant value of 1.4 for $p^*$ and 5/3 for 
$\Gamma_2$, which represent very well the $\delta$ Scuti regime.

When a range for the {\em phase~lag} and for $R$ is assumed a diagram
such as that shown in Fig~12 can be drawn, where amplitude ratios and
phase differences are calculated for the range $90\deg \leq \Psi^T \leq
140\deg$ and $0.25 \leq R \leq 1$ and for the discriminant color pairs
formed by $y$ and $b$ bands. In this review I will give the most usual
combination of a visual band, $y$, and a temperature indicator, $(b-y)$,
but other combinations are also useful, e.g. $v$ and $y$ as shown in
Garrido et al. (1990).  Uncertainties on the borders of these zones
can be of the order of $20\%$ because of the actual precision of model
atmospheres previously discussed. As can be seen in Fig~12 and originally
demonstrated by Stamford and Watson (1981), the regions of radial modes
are clearly separated from the regions of non-radial ones which in turn
can be even distinguished among them. The reason is that, in the $\delta$
Scuti regime, where the {\em phase~lags} are close to $100\deg$, a clear
phase shift is originated for different photometric bands for the lowest
$l$-values, as previously discussed. For $l$=3 the amplitude ratio is
very different from the other three $l$-values and this can be used to
discriminate it. These phase differences are however very small and
we need a very high precision photometry when determining phases and
amplitudes from a classical Fourier fitting to the time series. On the
other hand we do not know the {phase~lags} for the $\gamma$ Dor stars
and hence these diagrams are not very useful for them until reliable
values for these stars are known. At the end of this review I will give a
explanation on how to calculate $\Psi^T$ and $R$ values from multi-band
photometric observations and we will see that the values deduced for
the {\em phase~lags} for these variables are close to $0\deg$.

The effect of using different model atmospheres is not relevant, discrimination 
is also attained in a clear way and the only remarkable difference is the 
location of the photometric temperature indicator $(b-y)$, indicating a slight 
different temperature calibration for these two model atmospheres.   

\begin{figure}
\plotfiddle{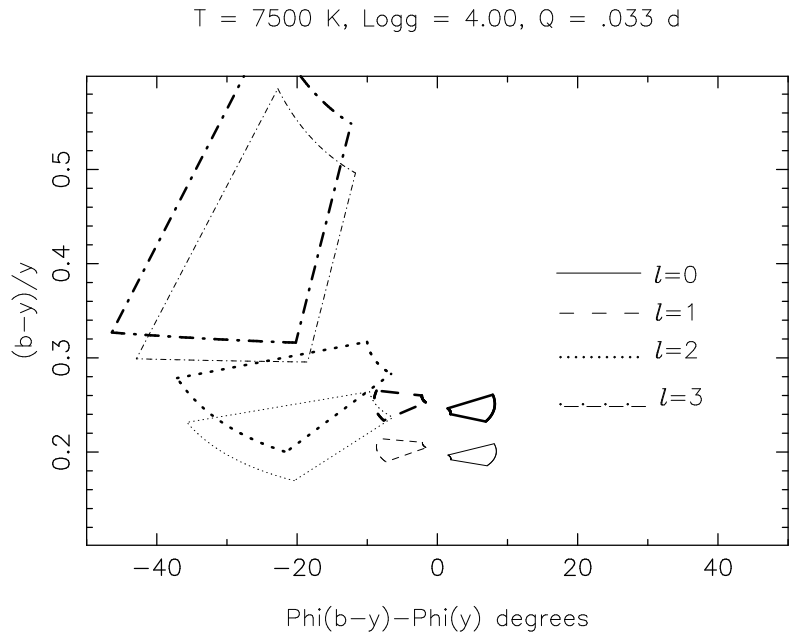}{8.5cm}{-90}{45}{45}{-180}{260}
\caption{``Regions of interest'' for a typical $\delta$ Scuti regime 
close to the fundamental radial $Q$-value of 0.033 days. Heavy lines are 
for the Kurucz models and light ones for S\&K models.
\label{fig-12}
}
\end{figure}

\begin{figure}
\plotfiddle{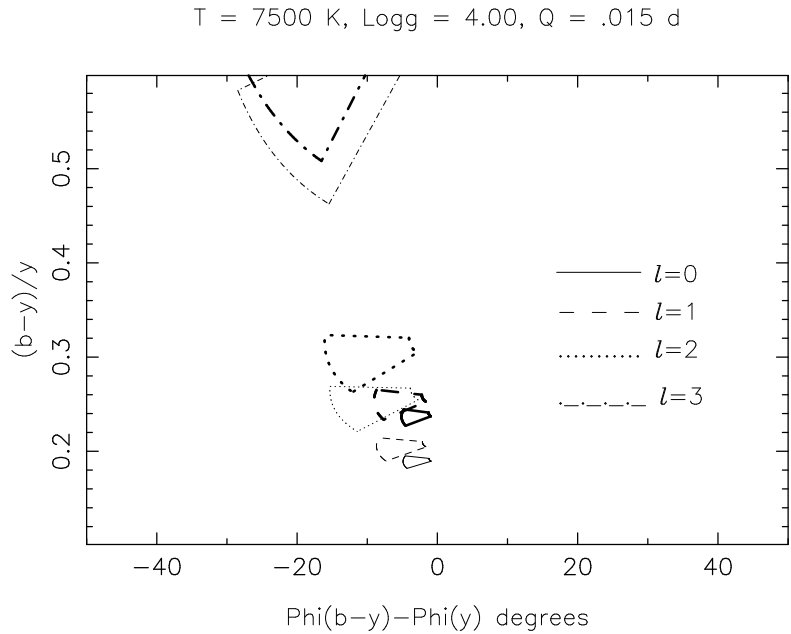}{8.5cm}{-90}{45}{45}{-180}{260}
\caption{Same as Fig~12 but for a $Q$-value of 0.015 days.
\label{fig-13}
}
\end{figure}

However the effect of the pulsation constant, is dramatic. When this value is 
lowered to a 
value of 0.015 days, corresponding to a radial 3rd or 4th overtone, then 
negative 
values are predicted also for radial modes and it is not true anymore that any 
negative value, in a $(b-y)$ vs. $y$ plot, corresponds to a non-radial mode. 
An example 
is given in Fig~13. 

When we go to higher temperatures the radial regions at positive phase 
differences begin to shrink and finally disappear at around 8500 K even for 
a radial value for $Q$. An intermediate case is plotted 
 in Fig~14 where the 
temperature is 8000 K and a higher radial value is assumed for $Q$. 
Notice the difference in amplitude ratios with respect to previous figures 
for the case 
$l$=3 which could be due to the effect of the temperature derivatives in the 
model atmospheres.   

\begin{figure}
\plotfiddle{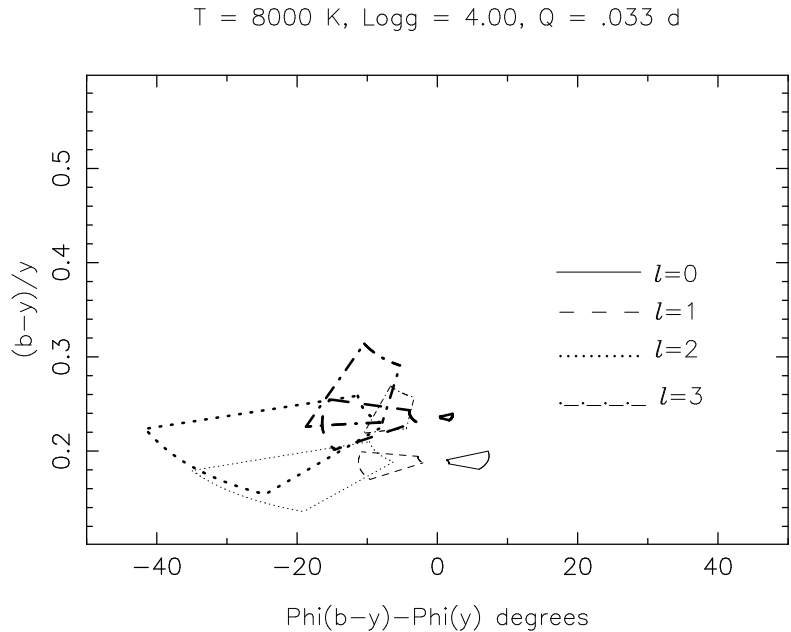}{8.5cm}{-90}{45}{45}{-180}{260}
\caption{Same as Fig~12 but for $T_{\rm eff}=8000~$K.
\label{fig-14}
}
\end{figure}

In conclusion, these ``regions of interest'' are very dependent on the chosen 
photometric bands and used model atmospheres. In 
particular, model atmospheres with smoother derivatives and without 
discontinuities are required. A good test for these model atmospheres would be 
to compare the observed photometric variations with theoretical 
 variations induced by a change in temperature. In other words not 
only to use standard stars for calibrating photometric indices but to use 
pulsating stars, with the above explained cautions, as 
standard stars regarding photometric variations.

$l$-values deduced from these considerations should be taken with
caution given the errors of the photometric measurements and the 
inconsistencies of
model atmospheres. However there is a way to alleviate this problem: to know 
the quantities $(\Psi^T, R)$ and the to solve equation (1) for $l$.


\section{Practical calculation of $(\Psi^T, R)$}

I review in this section the two 
 different methods to calculate actual values for the unknown quantities
 $(\Psi^T, R)$: A theoretical one given recently by Balona and Evers (1999) 
and another observational one given in Garrido et al. (1990).

\subsection{Theoretical approach}

Balona and Evers (1999) derive, from theoretical considerations, the
unknown quantities $(\Psi^T, R)$. Although they use $f$ instead of $R$
those parameters are related through the equation:

\begin{eqnarray}
 f & = & 4\,R\,C\,\frac{\Gamma_2 - 1}{\Gamma_2}
\end{eqnarray}

Unfortunately these parameters are very sensible to the treatment of the
convection specially for cool $\delta$ Scuti models, as shown by the
authors in their Fig~3. The onset of the convection occurs at $T_{\rm
eff} \leq 7900$~K; for higher temperatures the models are radiative
and $(\Psi^T, f)$ are constant and independent of the mixing length
parameter adopted.

 The authors develop a procedure which minimizes the distance between the 
observed and theoretically predicted color phases and amplitudes. These 
theoretical predictions come from non-adiabatic pulsation calculations of 
the couple $(\Psi^T, f)$ from a model which also predicts the observed 
frequency of the pulsating star. They give some identification for the best 
observed stars in the Str\"omgren multicolor photometric system in their 
Table~4.

 They apply the method to three double mode $\delta$ Scuti stars with period 
 ratios typical of radial pulsators, i.e. AE UMa: 0.773, BP Peg: 0.772 and
 RV Ari: 0.773, and they find good solutions in general but they note that 
discrimination between $l$=0 and $l$=1 is sometimes poor. For some other HADS 
they find no solutions and for another they have to select a $g$-mode as the 
best solution. The method can be limited by our ignorance of the convection 
but in principle could be very useful for radiative models, i.e. 
the hot regime of the $\delta$ Scuti stars.

\subsection{Observational approach}

This method was developed by Garrido et al. (1990) and uses multicolor 
photometry to derive the three unknown parameters $(\Psi^T, R)$ and $l$ in 
equation (1). 

At first sight a direct fitting to the formula appears to give very unstable 
values since its complexity prevents an inverse solution but, as indicated in 
that paper, if we assume no dependence on $\lambda$ for the limb darkening 
integrals, which is nearly true for the lowest $l$-values (see Fig~5, 6, 7 and 
8), equation (1) becomes reversible for any $l$-value. Errors are of the 
order of $1\%$ for $l$=1, $6\%$ for $l$=2, $18\%$ for $l$=3 and $30\%$ for 
$l$=4. In particular we 
need at least three bands, in order to have at least two color indices, but 
 in practice we have the four Str\"omgren bands so allowing to calculate 15 
different values for the couple $(\Psi^T, R)$.

Couples of values for $(\Psi^T, R)$ are plotted in Fig~15, for the
$\delta$ Scuti stars, where the error bars refers to 1-sigma value for
the 15 values calculated for each star of Table~1.  Errors bars for the
other values not plotted in Fig~15 are of the same order and are not shown
for clarity.  All the $R$ values fall in the range of $0.25 \leq R \leq
1$, which is just the expected range from theoretical arguments given
before. Also $80\deg\leq \Psi^T \leq~180\deg$ as indicated by the existing
simultaneous radial velocity observations. Values based on Kurucz models
are more concentrated than values for the S\&K models and the reason
is not clear. In order to see the effect of the large variations shown
by the $u$ derivatives in Fig~2, 3 and 4 I plot in them an average of
these derivatives over temperature and gravity. The result is a more
concentrated range for the $(\Psi^T, R)$ values, so indicating that
the dispersion shown by the derivatives could be a source of noise for
calculating these parameters, because of the non-smoothness of model
atmospheres.

\begin{figure}
\plotfiddle{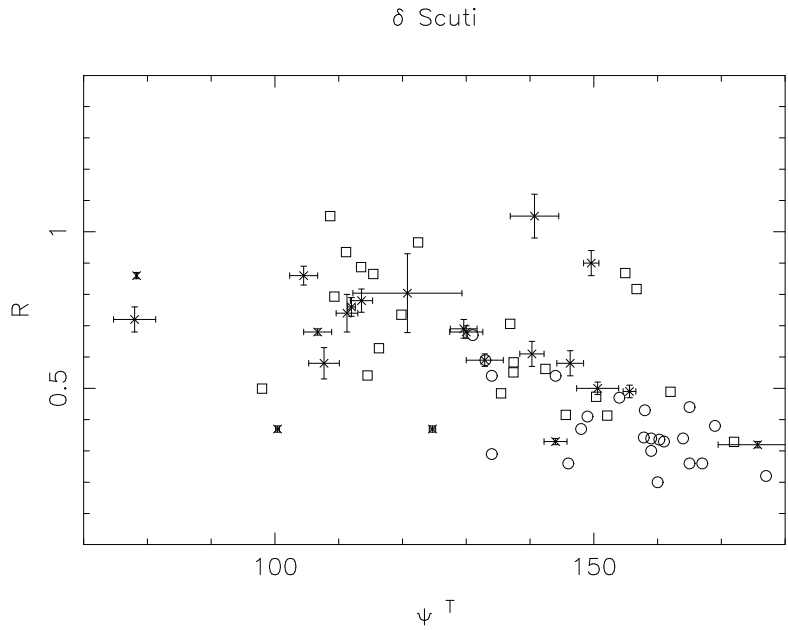}{7.5cm}{-90}{45}{45}{-180}{250}
\caption{{\em Phase lag} vs $R$ parameter for the $\delta$ Scuti stars
given in Table~1. Crosses are values derived from Smalley and Kupka
models, errors bars are 1-$\sigma$ standard deviations as explained
in the text, circles are for Kurucz models and squares are for average
values as explained in the text.  
\label{fig-15}
}
\end{figure}

\begin{figure}
\plotfiddle{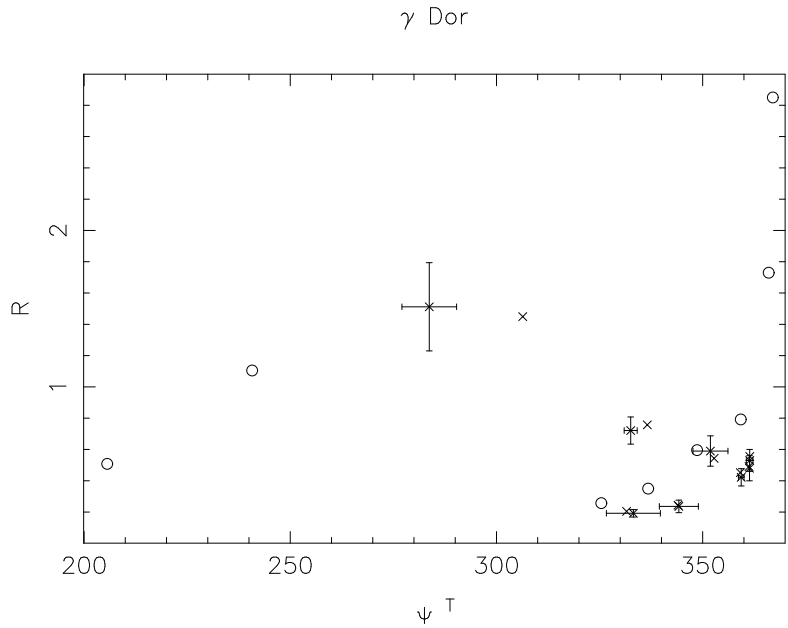}{8.5cm}{-90}{45}{45}{-180}{260}
\caption{{\em Phase~lag} vs $R$ parameter for the $\gamma$ Dor stars given in 
Table~2. Symbols are the same as in Fig~15.
\label{fig-16}
}
\end{figure}

\begin{figure}
\plotfiddle{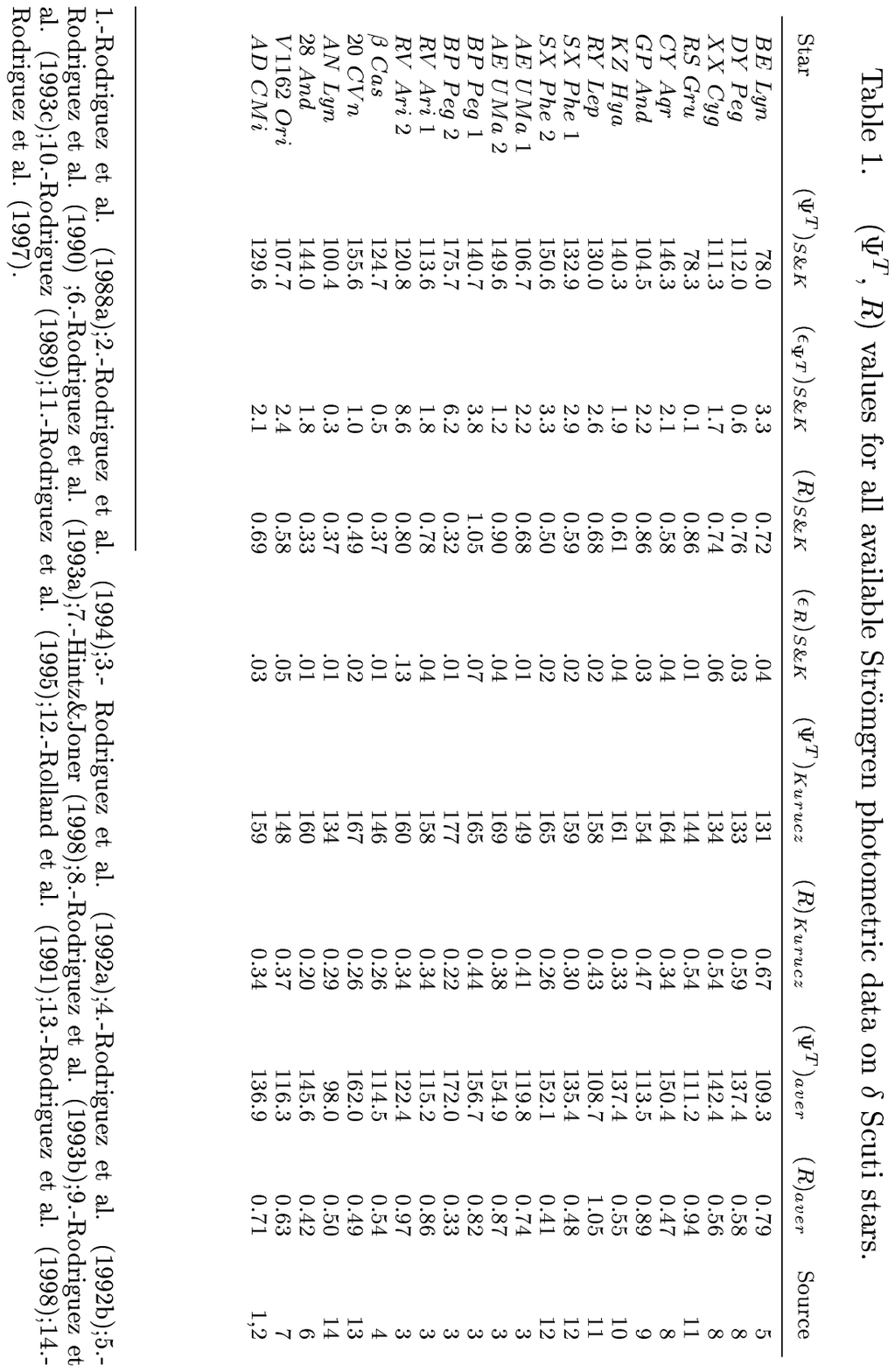}{15cm}{180}{120}{120}{390}{805}
\end{figure}

\begin{figure}
\plotfiddle{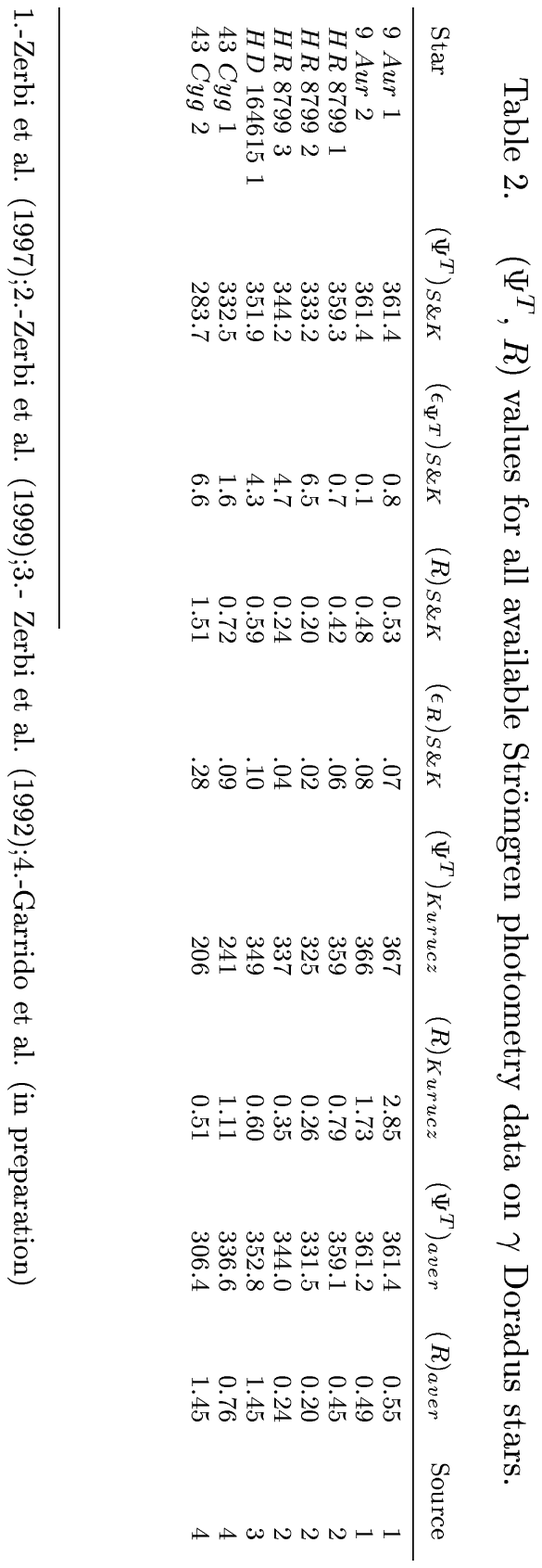}{15cm}{0}{120}{120}{-475}{-385}
\end{figure}

In Fig~16 the couples $(\Psi^T,R)$ are plotted for the $\gamma$
Dor stars.  These values are more spread basically for two reason: the
small amplitude of the variations for these stars and the small phase
differences observed for the photometric indices. Some physical unfounded
$R$ values, i.e. greater than unity, appear for some stars. In any case
the values derived using the Kurucz models show larger dispersions. This
fact seems to indicate that Kurucz models are less stable for lower
temperatures, since $\gamma$ Dor stars are cooler than the majority of the
$\delta$ Scuti stars. This fact could be also the explanation of the more
concentrated values found before for $\delta$ Scuti stars. In any case the
phase differences observed in the Str\"omgren bands for these stars seem
to indicate that {\em phase~lags} for these stars  are close to zero. This
is the first time that these values are calculated for these stars and,
if confirmed, could be an important clue to determine the pulsational
characteristics of these not very well understood variables. It is to
be noted here that $\Psi^T=180\deg$ for an adiabatic atmosphere and that
$\Psi^T=0\deg$ for a stellar spot!

\section{Some examples of practical identifications}

\subsection{$\delta$ Scuti stars}

The next step following the flow chart of Fig~1 is to calculate, from equation 
(1) and with known {\em phase~lags} and $R$ known, the $l$-value which minimizes 
the distances from the predicted values to the observed ones in all the 
photometric bands. To do that I have considered phases and amplitudes 
separately.

The best fit, in the sense of minimum variance for amplitudes, is 
marked ``Amplitude variance'' in the plots 17, 18, 19 and 10 and the 
corresponding to the phases, is 
marked ``Phase variance''. The reason is that when looking at the ``regions 
of interest'' there are some $l$-values which are separated in the phase 
difference axis, like $l$=0 and 1 in Fig~12 and other are separated in the 
amplitude ratio axis, like $l$=3. 

As already mentioned the best way to test these identifications is to
apply the method to the well known double mode $\delta$ Scuti stars which
are oscillating into two radial modes. In Fig~17 I plot the results for
these stars for the first four $l$-values and for the two comparison
axis: amplitudes and phases. As can be seen all the minima, both for
fundamental and overtone modes, in the ``phase variance'' panels, are
located always and in a very clear manner, except maybe for the first
overtone of BP~Peg, in the $l$=0 position as expected from its radial
nature. However the minima regarding ``amplitude variance'' are not very
different for the first $l$=0, 1, 2 values reflecting the behavior seen
in Fig~12, in the sense that phases are discriminant in this regime for
the lowest $l$-values and amplitudes are for the higher ones. A blind
application of the method for phases and amplitudes together could
introduce then some extra variance which could change the minimum. It
is important to know, before doing these comparisons, where the star
regime falls in the phase difference vs amplitude ratio diagram.

When the method is applied to HADS the results are clear: all of them are 
pulsating in a radial mode (Fig~18).

\begin{figure}
\plotfiddle{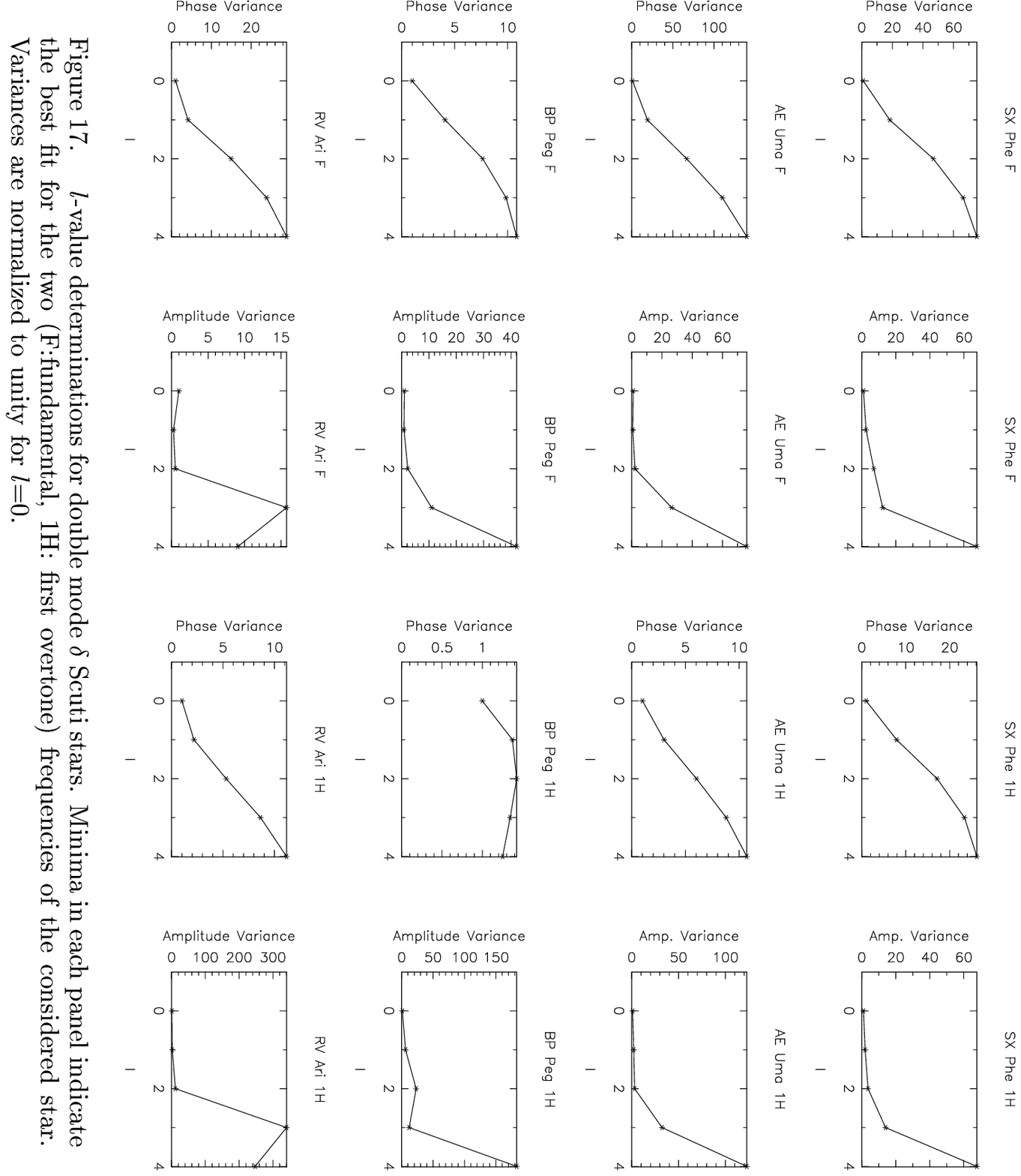}{15cm}{0}{90}{90}{-205}{-270}
\end{figure}
\addtocounter{figure}{1}

\begin{figure}
\plotfiddle{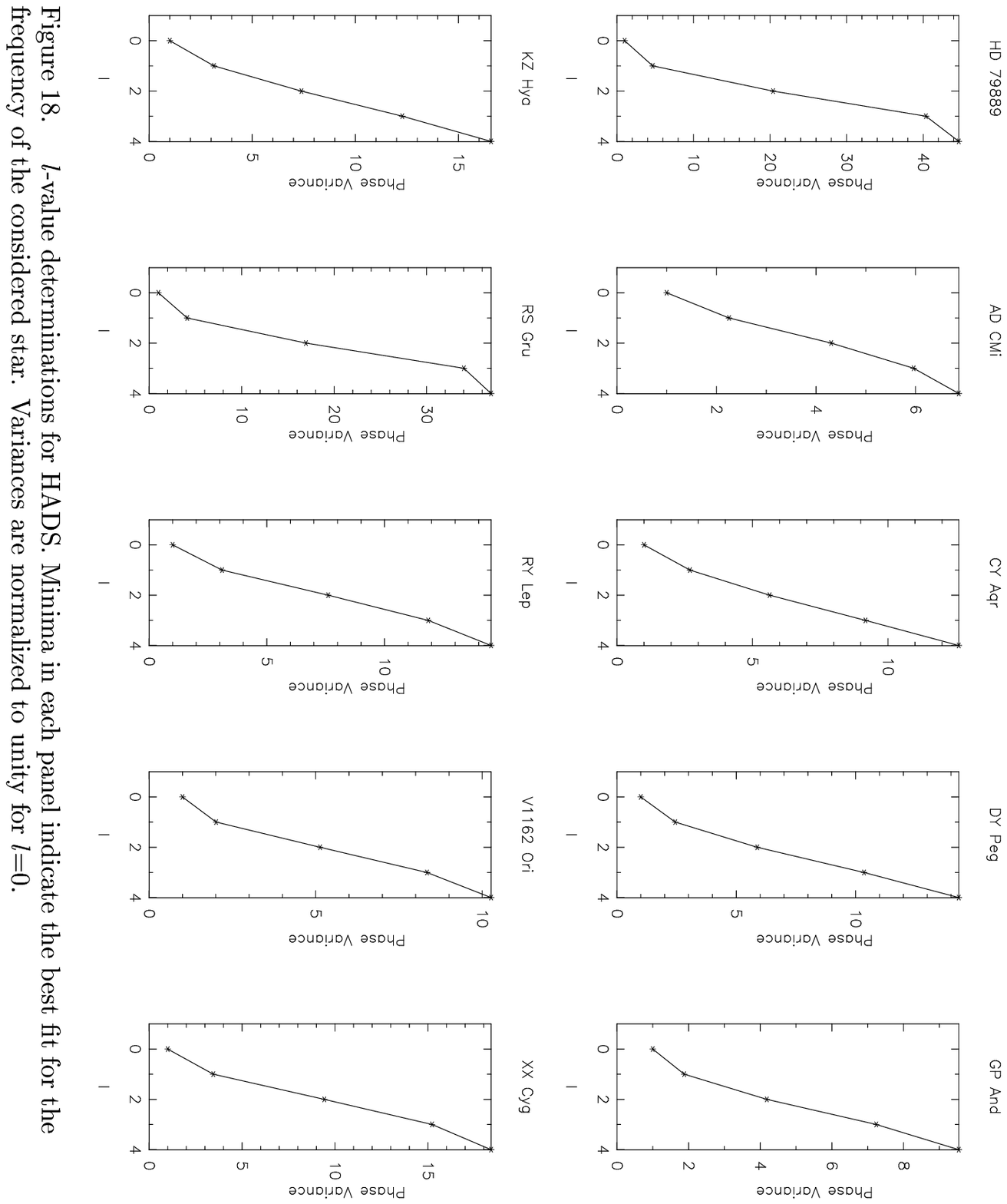}{15cm}{180}{90}{90}{205}{720}
\end{figure}
\addtocounter{figure}{1}

Nevertheless when applied to the low amplitude $\delta$ Scuti stars
the results are less conclusive because of the small amplitudes and the
corresponding larger relative errors in the determination of amplitudes
and especially phases. The results for the three best studied stars
(due to their monoperiodicity) plotted seem to indicate that at least
20 CVn is oscillating in a radial mode, but the other two seem to be
pulsating in an $l$=1 or 2 mode (Fig~19).

\begin{figure} \plotfiddle{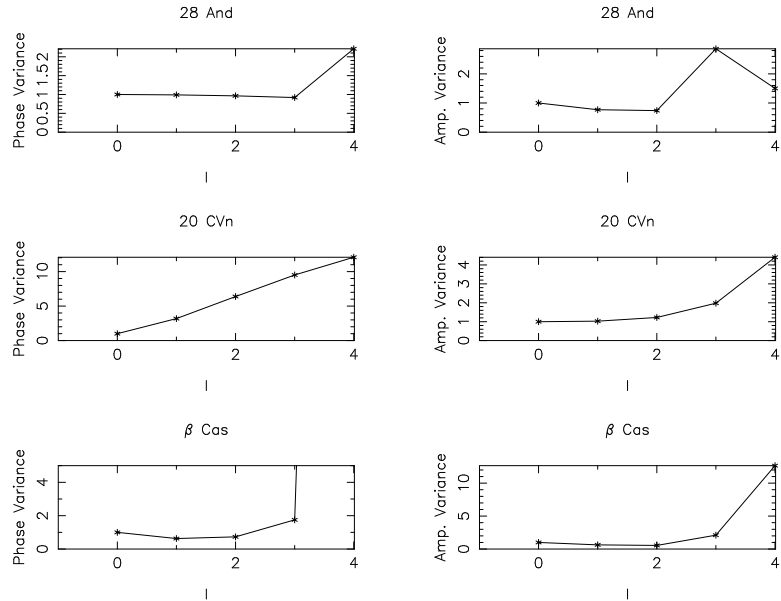}{8cm}{-90}{42}{42}{-160}{240}
\caption{$l$-value determinations for low monoperiodic $\delta$ Scuti
stars.  Minima in each panel indicate the best fit for the frequency
of the considered star. Variances are normalized to unity for $l$=0.
\label{fig-19} } \end{figure}

\subsection{$\gamma$ Doradus stars}

The results for the $\gamma$ Dor stars, as indicated in Fig~20, show that
all frequencies for all stars seem to be $l$=1 non-radial modes if we
accept the ``phase variance'' indications, whereas the discrimination
is not so clear for the ``amplitude variance'' plots, indicating in
this case the presence of modes with $l$=1, 2 but not certainly 3. As
explained in the previous section it would be convenient, in order to
select the most relevant discriminator, to construct a ``region of
interest'' for these variables with the new {\em phase~lag} derived
before. Such a plot is shown in Fig~21, where the relevant zones have
been calculated for {\em phase~lags} close to zero and atmospheric
characteristics different to those of a $\delta$ Scuti star plus a very
different pulsation constant which for a $\gamma$ Dor star is an order
of magnitude larger. It is clear from that figure that discrimination
between $l$=1 and $l$=2 modes is based on phase differences and they
are not distinguishable from amplitude ratios; $l$=3 is however well
discriminated from the other two lower $l$-values. Going back to the
interpretation of the results shown in Fig~20 for $\gamma$ Dor stars,
it becomes clear from the amplitude diagrams that these stars are not
oscillating in an $l$=3 mode and, from the phase diagrams, that they
are probably oscillating in an $l$=1 mode. However the small amplitudes
and small phase differences observed in these stars prevent me to be
sure about these conclusions. It would be very interesting to perform
simultaneous photometry and spectroscopy for these objects in order to
have real measurements of their {\em phase~lags}.

\begin{figure}
\plotfiddle{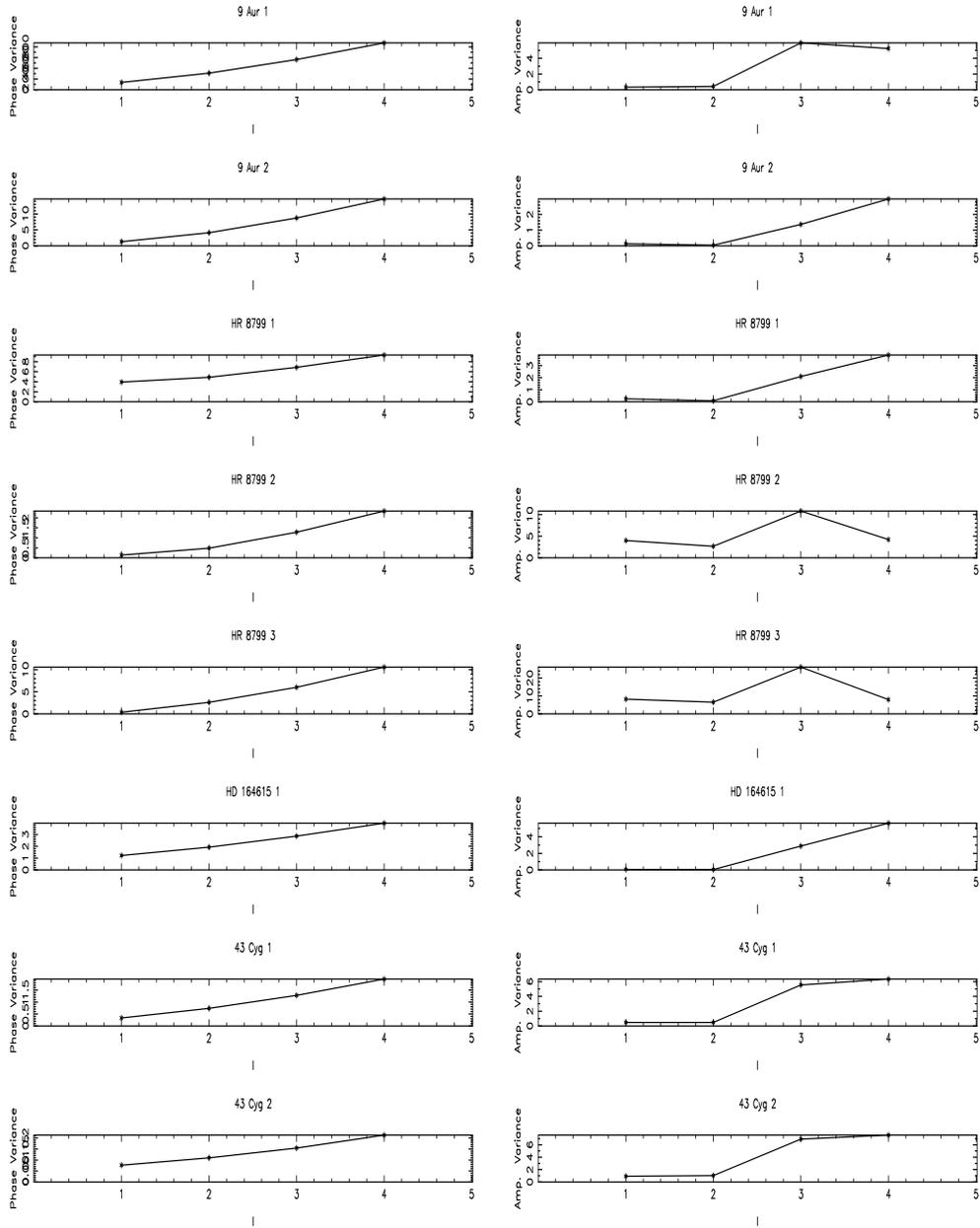}{17cm}{-90}{51}{85}{-198}{510}
\caption{$l$-value determinations for $\gamma$ Dor stars. 
Minima in each panel indicate the best fit for the frequency of the 
considered star. Variances are normalized to unity for $l$=0. Left panels for 
``phase variance'' and right panels for ``amplitude variance''.  
\label{fig-20}
}
\end{figure}

\begin{figure}[t]
\plotfiddle{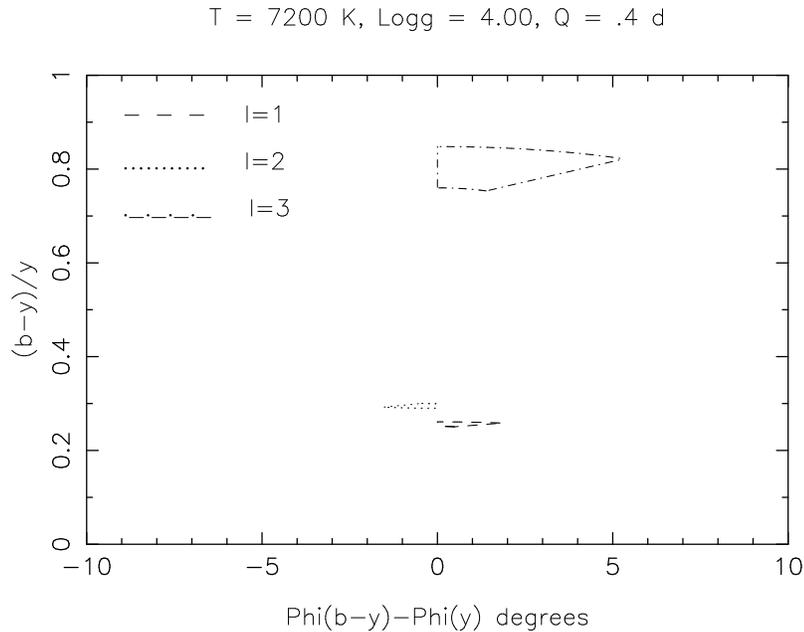}{8.5cm}{-90}{45}{45}{-180}{260}
\caption{``Regions of interest'' for a $\gamma$ Dor star with $320\deg~
\leq~\Psi^T~\leq~360\deg$ and $0.25~\leq~R~\leq~1$. Please note the different 
atmospheric characteristics with respect to a $\delta$ Scuti star and the very 
different pulsation constant. 
\label{fig-21}
}
\end{figure}

\section{Color information in space missions}

The motivation to do asteroseismology from space is to detect and analyze
solar like oscillations and to learn about the stellar interiors in
other stars than the Sun. Color informations for these observations are
not relevant since the modes can be easily identified from the rotational
splitting.  Due to the extremely small amplitudes expected to be measured,
the number of photons is a crucial quantity in order to decrease the noise
and therefore white light is the preferred solution. Modal identification
is made by direct measuring of the rotational splitting. However it
was previously demonstrated that multicolor information is essential to
identify the modes for the stars we are interested in this conference,
i.e.  $\delta$ Scuti and $\gamma$ Dor variables.  Asteroseismological
techniques become therefore possible opening the possibility to test
stellar interiors in this region of the HR diagram.

COROT {\it (http://www.astrsp-mrs.fr/www/ecorot.html)} is a French
asteroseismological and planet detection space mission which is now under
study by the CNES.  The scientific mission is explained in another part
of this meeting. Basically there are two CCDs in the focal plane, one of
them dedicated to asteroseismology, mainly for solar like oscillations,
with no color information, and the other dedicated to planet detection
by transit with color information in order to distinguish transits from
other active phenomena in stars. This color information is achieved
through the interposition of a prism in front of the CCD. The idea is
to download two or three parts of this spectrum so giving different
photometric colors. The field will contain several thousands of stars in
order to be able to detect some events due to terrestrial planet transits
during the two years for which the mission is expected to be operative.

MONS is mission lead by the Danish community which is also under study
and will probably offer some color information. The reader is referred to
another part of this meeting. Although the mission is not yet decided,
it would probably have a dichroic prism in order to measure simultaneously
two colors.

In any case these missions will provide data with very high precision and 
then much of the problems we have now for extracting information from 
photometry probably will disappear. We will be able to get 
data for multiperiodic $\delta$ Scuti stars and also for $\gamma$
 Dor with large enough amplitude to make feasible the photometric techniques 
that I review in this report. 

I have made some simulations, constructing an arbitrary 12 color
photometric system, for these two kind of pulsating stars in order to see
the order of magnitude of the phases differences and amplitude ratios
among the different photometric bands. These bands have arbitrarily
100~\AA\ width distributed over the visible spectrum each 300~\AA.
I present the results in Fig~22 for a $\delta$ Scuti regime and in
Fig~23 for a $\gamma$ Dor regime.  As shown there the main discriminant
factor for radial and non-radial modes in $\delta$ Scuti stars is the
phase difference between two colors, being also possible, depending
on the precision, to discriminate among the non-radial modes. The main
contribution of the amplitude ratios is to separate the high order $l$=4
mode from the others. In any case a combination of the two diagrams will
provide, at the precision we need, a clear identification for the lowest
$l$-values in the sense that the larger the difference in wavelength
the better the discrimination, at least up to 3700~\AA.

\begin{figure}[t]
\plotfiddle{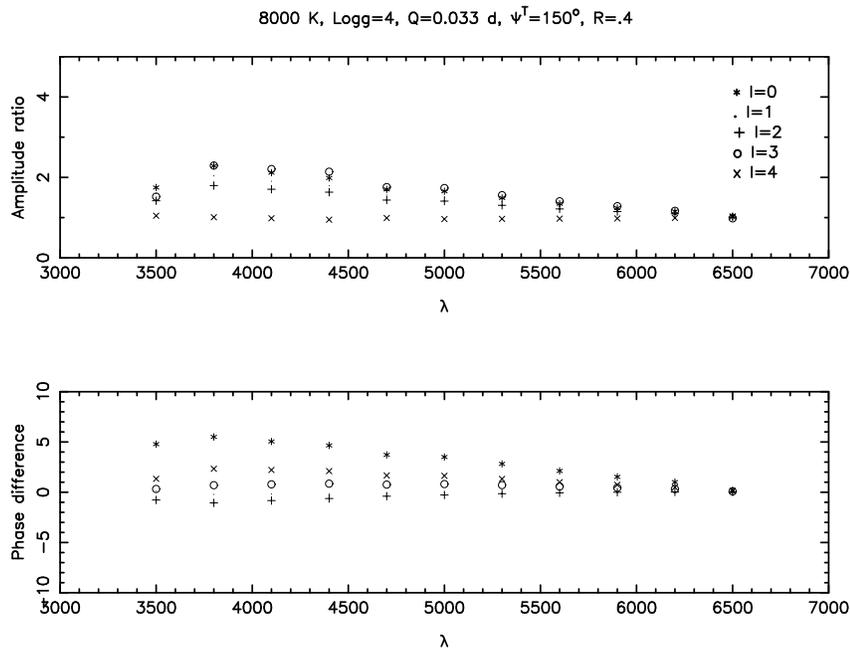}{8.5cm}{-90}{45}{45}{-180}{260}
\caption{Amplitude ratios and phase differences for a 12 color photometric 
system in the visible range for a typical $\delta$ Scuti regime. Ratios and 
differences are relative to the photometric band centered at 7000~\AA.
\label{fig-22}
}
\end{figure}

\begin{figure}[t]
\plotfiddle{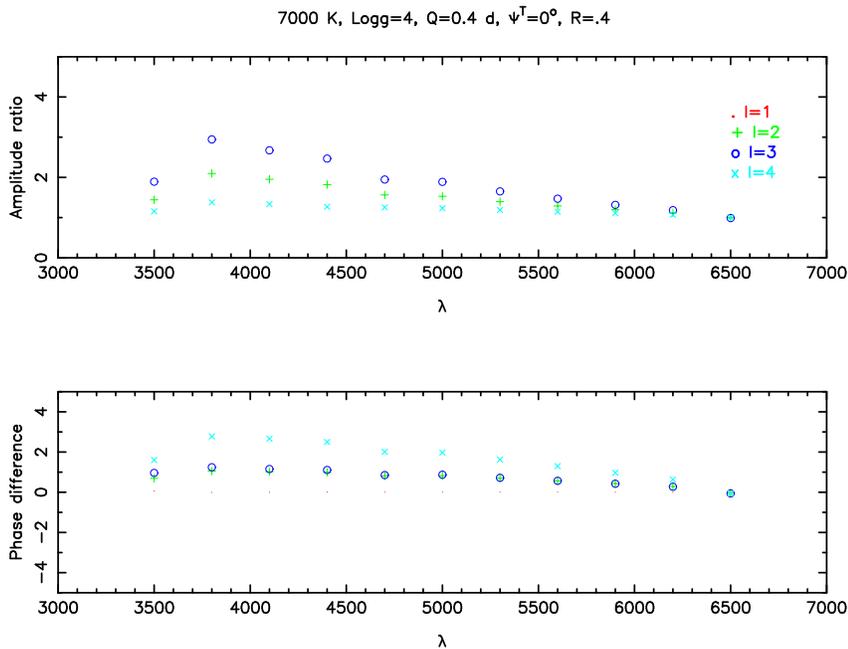}{8.5cm}{-90}{45}{45}{-180}{260}
\caption{Amplitude ratios and phase differences for a 12 color photometric 
system in the visible range for a typical $\gamma$ Dor regime. Ratios and 
differences are relative to the photometric band centered at 7000~\AA.
\label{fig-23}
}
\end{figure}

For $\gamma$ Dor stars the situation is slightly different mainly due the the 
very high pulsation constant and the very different {\em phase~lag}. As shown 
before the amplitude ratios do not separate $l$=1 from $l$=2 but the other 
non-radial modes are more clearly discriminated. The same is true for $l$=2 
and $l$=3 concerning phase differences but fortunately the other modes can be 
separated. The combined information from the two 
panels could be of capital importance to separate modes.

With the precision we expect to have in COROT, from 100 to 10 ppm
depending on the magnitude in the exoplanet focal plane, I estimate
here phases and amplitudes will be so small that the method described
becomes realistically applicable.  I think that under these conditions
an asteroseismological study of these variables is possible without the
ambiguities of miss-identifications
 of the modes.

\section{Conclusions}

 We already know that simple matching of observed to theoretically 
calculated frequency 
spectra for the best observed $\delta$ Scuti variables, such as FG Vir in 
Breger et al. (1999), do not give a unique solution. Possible combinations 
varying
 some physical inputs of the static models do not allow us to constraint the 
theoretical model. An identification procedure is then required for the 
observed modes. 

We have seen that the success of the photometric method to identify 
 the degree of the spherical harmonic $l$ for $\delta$ Scuti and $\gamma$ Dor 
stars depends on the model atmospheres we are using and on the precision on 
the photometric data we are analysing.

Although the global uncertainties bound to the photometric 
calibrations, i.e. calculation of $T_{\rm eff},~\log\,{\sl g},~[M/H]~$and$~Q$, do not 
affect 
dramatically to the discrimination procedure, the model atmospheres present 
some subtle characteristics. 
 Kurucz models present some inconsistencies, at the level of 
the required continuities in the fluxes and derivatives which, in the HR 
region where the stars we are investigating fall, seem to be related with 
the treatment of convection and in particular of overshooting. Other 
improved 
models, such as those described in Smalley and Kupka (1997), present a  
smoother behavior at these temperatures and gravities but some inconsistencies 
still remain regarding continuities in the flux derivatives. An  
appropriate comparison with very accurate photometric data on the variables 
could be useful to improve these discontinuities in the models.

Photometric precision is now sufficient only for the highest amplitude
pulsating variables. In particular, nowadays existing HADS photometric
data with precisions of around 1 mmag, are useful to identify their
oscillation modes.  From observations of other variables with lower
amplitudes I estimate that we need at least signal to noise ratios of
the order of 20--30 in order to have a stable Fourier solution and then
make the photometric method feasible.

The method supplies reasonable $l$-values for HADS, i.e. all of them are
found to be radial pulsators, and for some other $\delta$ Scuti stars
with lower amplitudes non-radial modes seem to be identified. In any
case when applied to known radial pulsators, as the double mode $\delta$
Scuti stars, the results are consistent with their radial nature.

When applied to the new discovered $\gamma$ Dor stars the method is able
to give estimations of two not very well known physical quantities,
$\Psi^T$ and $R$. The {\em phase~lags} for these stars, if confirmed,
would be very different from the classical value observed in $\delta$
Scuti stars. Furthermore the existing multi-band photometry for some of
them seems to indicate that these stars are very probably oscillating
in an $l$=1 mode.

It is also shown that, within the uncertainties, the method provides
estimations of $(\Psi^T,R)$ which can be used to compare with theoretical
models.  These quantities depend very much on the convection parameters
which are not very well known and could be useful to modelize the
convection in these stars, which in turn could give important clues to
the understanding of the onset of the convection in this region of the
HR diagram.

The new generation of asteroseismological space missions will provide in
a next future a huge quantity of very high quality data for these stars.
As demonstrated in this review and in order to understand and disentangle
the complicated frequency spectrum of multiperiodic $\delta$ Scuti stars
we need colored information as demonstrated in this review.

Actual model atmospheres need to be improved, specially the smoothness in 
the derivatives and limb darkening variations, in order to extract the whole 
potential of the photometric discrimination method.

\acknowledgments

I am very grateful to Enrique Solano for supplying me the Kurucz fluxes 
and limb darkening coefficients and to Antonio Claret for the PHOENIX limb 
darkening coefficients. I am also grateful to Chris Sterken for many helpful 
comments, and to the editors for making possible a more readable
version of the manuscript.

%

\end{document}